\documentclass[journal]{IEEEtran}
\IEEEoverridecommandlockouts
\usepackage{booktabs}
\usepackage{cite}
\usepackage{graphicx}
\usepackage{subcaption}
\usepackage[cmex10]{amsmath}
\usepackage{algorithmicx}
\usepackage[ruled]{algorithm}
\usepackage{algpseudocode}
\usepackage{mathtools}
\DeclarePairedDelimiter\abs{\lvert}{\rvert}

\usepackage{amsmath}
\usepackage{amsfonts}
\usepackage{amssymb}
\DeclareMathOperator*{\argmax}{argmax}
\usepackage{multirow}
\usepackage{textcomp}
\usepackage{array}
\usepackage{fixltx2e}
\usepackage{stfloats}
\usepackage{url}

\def\BibTeX{{\rm B\kern-.05em{\sc i\kern-.025em b}\kern-.08em
    T\kern-.1667em\lower.7ex\hbox{E}\kern-.125emX}}

\begin{document}

\title{Real-Time Video Content Popularity Detection Based on Mean Change Point Analysis}

\author{Sotiris~Skaperas,~\IEEEmembership{Student Member,~IEEE,}
        Lefteris~Mamatas,~\IEEEmembership{Member,~IEEE,}
        and~Arsenia~Chorti,~\IEEEmembership{Member,~IEEE}
\thanks{Sotiris Skaperas and Lefteris Mamatas  are with the Department of Applied Informatics,
University of Macedonia, Greece, \{sotskap,
emamatas\}@uom.edu.gr}
\thanks{Arsenia Chorti is with ETIS / Universit\'e Paris Seine, Universit\'e Cergy-Pointoise, ENSEA, CNRS, France,
arsenia.chorti@ensea.fr.}
\thanks{Manuscript received July 2019.}
}


\maketitle

\begin{abstract}
Video content is responsible for more than $70\%$ of the global IP traffic. Consequently, it is important for content delivery infrastructures to rapidly detect and respond to changes in content popularity dynamics. In this paper, we propose the employment of on-line change point (CP) analysis to implement real-time, autonomous and low-complexity video content popularity detection. Our proposal, denoted as \textit{real-time change point detector (RCPD)}, estimates the existence, the number and the direction of changes on the average number of video visits by combining: (i) off-line and on-line CP detection algorithms; (ii) an improved time-series segmentation heuristic for the reliable detection of multiple CPs; and (iii) two algorithms for the identification of the direction of changes. The proposed detector is validated against synthetic data, as well as a large database of real YouTube video visits. It is demonstrated that the RCPD can accurately identify changes in the average content popularity and the direction of change. In particular, the success rate of the RCPD over synthetic data is shown to exceed $94\%$ for medium and large changes in content popularity. Additionally, the dynamic time warping distance, between the actual and the estimated changes, has been found to range between $20$ samples on average, over synthetic data, to $52$ samples, in real data. The rapid responsiveness of the RCPD is instrumental in the deployment of real-time, lightweight load balancing solutions, as shown in a real example.
\end{abstract}

\begin{IEEEkeywords}
Video content popularity detection, change point analysis, on-line change point detection, binary segmentation algorithm, load balancing.
%
%
%
\end{IEEEkeywords}

%
\IEEEpeerreviewmaketitle

\section{Introduction}

\IEEEPARstart{V}{ideo} content is projected to account for $82\%$ of the global Internet traffic by 2020, significantly increased from $72\%$ in 2016 \cite{ersi-phd-1}. In parallel, novel emerging networking, cloud and edge computing paradigms with significant elasticity capabilities appeared recently, e.g., software-defined networks (SDN) \cite{openflow}, cloud orchestration proposals \cite{necos} and content distribution networks (CDNs) \cite{chronis-catania}. These advances offer the means to respond quickly to changes in content popularity dynamics with appropriate adaptations, e.g., in terms of efficient server resource allocation schemes, load balancing or content caching. As a result, the early detection of changes in content popularity \cite{springer-survey}, \cite{survey-30} is proving a highly important topic and can have a significant impact on the network traffic and the utilization of servers.

So far, the vast majority of research efforts have focused on the \textit{prediction} of content popularity dynamics, as opposed to their \textit{real time detection}, which is the focus of this study. There is a multitude of reasons as to why the precision of even state-of-the-art prediction algorithms can be impaired. A variety of factors -- both from the digital and the physical world -- can influence the users' Internet surfing behavior, e.g., \cite{springer-survey}: (i) the quality, type (e.g., commercial or user-provided) and life-time of content; (ii) its relevance to users and physical events; (iii) the social interactions between users; and (iv) the content promotion strategies involved. Importantly, mid-term and long-term content popularity prediction \cite{survey-32} -- and corresponding adaptations in the network or cloud environment -- can prove highly inaccurate \cite{cascade} and thus result in sub-optimal service planning, provisioning, and utilization of resources or violation of service level agreements. 

In this work, to address the aforementioned shortcomings of the commonly employed prediction algorithms, we propose a corresponding detector, referred to as the ``real-time change point detector'' (RCPD). The RCPD is compatible with modern, flexible networking and cloud approaches, that are highly adaptive and can respond to short-term network dynamics. With accurate, on-line content popularity detection, discrepancies between inaccurate predictions and actual changes can be alleviated. The RCPD is real-time, lightweight, accurate and is parameterized autonomously by analyzing historical data. 

\par In the RCPD, we employ the change point (CP) detection theory and algorithms; their suitability is confirmed against a large number of synthetic as well as real YouTube video datasets. In this contribution, the early detection of changes in the average content popularity is addressed with a novel CP detection methodology, consisting of a training phase, using historical data, and, an on-line phase. In the training phase, we employ a modified off-line CP detection scheme to configure the on-line (sequential) algorithm's parameters. This approach is shown to greatly improve the accuracy of the on-line detector, as in essence, the algorithm parameterization is not arbitrary but rather extracted from corresponding historical data. To the best of our knowledge, it is the first time in the literature that retrospective (off-line) and sequential (on-line) CP detection schemes are combined in a single algorithm operating autonomously (i.e., without manual configuration of parameters).  


\par Besides that, our approach complements the off-line scheme with an improved time-series segmentation heuristic for the detection of multiple CPs.  Furthermore, we propose two possible variations for the on-line CP algorithm, the first based on the standard cumulative sum (CUSUM) procedure \cite{b10} and the second on the ratio-type CUSUM procedure \cite{b11}\footnote{The advantage of ratio-type CUSUM is that it does not require the estimation of long-run covariance (variance) matrices, which is the case for the standard CUSUM method.}. Additionally, we introduce two alternative indicators to detect the direction of changes: the first one is directly derived from the statistical test of the on-line CP procedure, while the second is based on a modified exponential moving average filter, extensively used in econometrics. As discussed in Sections III and IV, the RCPD combines all the above mentioned algorithmic elements, and is based on sufficiently general and convenient assumptions. Moreover, unlike other approaches e.g., \cite{brodsky}, we employ methods that allow dependence between observations (in the form of $t-$dependence), leading to more realistic assumptions for the statistical structure of the content visits. 

We evaluate the proposed detector and its individual algorithmic components (i.e., the off-line / on-line test statistics, the time-series segmentation algorithm and the trend indicator), over synthetic and real YouTube content views data. Our experiments using synthetic data, generated by an autoregressive moving average (ARMA) filter, demonstrate: 
\begin{itemize}
\item The superior performance of the proposed time-series segmentation heuristic over the standard approach, improving the true alarm rates by up to $43\%$. 
\item The ability of the two proposed trend indicators to identify the direction of estimated changes, with successful identification rates exceeding $99\%$, in all cases.
\item The RCPD performance; the true alarm rates surpass $94\%$ for medium / large changes in the mean number of content views, while the corresponding CP identification lag ranges between $10$ to $20$ instances, confirming the real-time operation of the detector. On the other hand, the RCPD achieves very small false alarm rates, well within the limits of the statistical error specified by the chosen significance level of the CP algorithms.
\end{itemize}

Furthermore, our tests on real YouTube content views datasets show that:
\begin{itemize}
\item YouTube video views match the underlying assumptions of the RCPD, i.e., the content popularity time-series datasets can be modeled as $t$-dependent.
\item The RCPD can detect CPs in more than $70\%$ of the videos in our dataset, implying a sufficiently high number of content popularity changes and the suitability of the CP theory framework for content popularity detection. 
\item The successful CP direction identifications exceed $91\%$, i.e., the proposed trend indicators work for real data. 
\item The average dynamic time warping (DTW) distance \cite{dtw}, \cite{dtw2} between the identified CPs and a benchmark off-line algorithm was estimated to be $52$ time instances on average, showcasing the rapid responsiveness of the RCPD. 
\item The overall processing cost of the RCPD is very low; notably, it took less than one second to process $882$ videos on a typical personal computer (PC).
\end{itemize}

Finally, as a proof-of-concept, we demonstrate the applicability of the proposed algorithm in a real load balancing scenario. We provide a set of measurements showcasing improvements in terms of the clients' connectivity time to download specific content, without a significant impact on the utilization of the content servers. This is achieved due to the deployment of additional content caches, an event triggered by the output of the proposed RCPD detector. 

The rest of the paper is organized as follows. In Section II, we discuss our approach with respect to related works. In Section III, we present the training phase of the RCPD algorithm, while the on-line phase is discussed in Section IV. In Section V, we present four experiments over synthetic data, providing an extensive validation of the RCPD and its subroutines, while in Section VI, we discuss corresponding experiments using a database of real YouTube video views. In Section VII, we demonstrate the load balancing gains achieved through the use of the RCPD, in a realistic content provisioning scenario. Our conclusions and directions for future work are presented in Section VIII.

\section{Related Works}

\par In this Section, we discuss how this work relates to the literature of video content popularity prediction, on one hand, and, anomaly detection (i.e., CP analysis), on the other hand.

The topic of content popularity attracted a lot of attention in recent years, because of its importance in a number of applications, such as network dimensioning (e.g., capacity planning or scaling of resources), on-line marketing (e.g., advertising, recommendation systems) or real-world outcome prediction (e.g., analysis of economical trends) \cite{springer-survey}. The main approaches used for content popularity estimation can be categorized as: (i) cumulative growth studies, estimating the ``amount of attention" from the publication instance to the prediction moment  \cite{survey-30}; (ii) temporal analysis approaches, i.e., how content visits evolve over time \cite{survey-33}; and (iii) clustering methods of content with similar popularity trends \cite{survey-32}. We note that many content popularity studies consider the aggregate behavior of a particular content, e.g., \cite{survey-30}, \cite{survey-33}, whereas we study the real-time behavior of video views time-series. In addition, studies using clustering methods \cite{survey-32} are based on content popularity prediction and adopt parametric models, unlike the RCPD algorithm that is non-parametric.  


\par To the best of our knowledge, our earlier conference paper \cite{GLOBECOM} is the first in the literature proposing CP techniques \cite{basev} for content popularity detection. The RCPD algorithm falls into the general category of anomaly detection \cite{chandola}; in essence, we assume that no changes in popularity constitutes the normal behavior of video content and search for deviations from this behavior. Non-parametric anomaly detection has typically been considered for the detection of abnormalities in the network traffic. As an example, in \cite{tarta13} an algorithm was proposed based on the Shiryaev-Roberts procedure for anomaly detection in computer network traffic. In \cite{tarta06} and \cite{wang}, CUSUM based approaches were introduced for the detection of SYN attacks. 

Further examples of parametric anomaly detection methods include \cite{thatte}, in which a bivariate sequential generalized likelihood ratio test (LRT) was proposed, accounting for the packet rate -- assumed to follow a Poisson distribution -- and the packet size  -- assumed to follow a normal distribution. Other parametric anomaly detection approaches assume a particular underlying process for the normal behavior and search for anomalies on the residuals of the process. For example, in \cite{soule}, Kalman filtering is combined with several CP methods, such as CUSUM and LRT, to detect anomalies in origin-destination flows. In \cite{noda}, traffic flows (in the form of TCP's finite state machine), are modeled using Markov chains and an anomaly detection mechanism based on the generalized LRT algorithm is developed. 
\par As opposed to previous content popularity prediction works, in this paper we introduce a novel CP detection methodology that provides accurate, lightweight, autonomous and on-line CP detection of content popularity. We formulate the detection of a change in the average content popularity as a statistical hypothesis test and employ non-parametric procedures to avoid a particular distribution assumption (such as a specific copula model). This context ensures low convergence time since it avoids estimating a large number of model parameters and restrictive assumptions that may not match the structure of the time-series. Furthermore, we avoid problems of parametric models that require parameters' fitting and selection, which become challenging as new data become available. In the proposed RCPD algorithm, an off-line phase specifies important parameters for the on-line phase; these parameters are re-evaluated dynamically after a detected CP. Our load-balancing experiments, elaborated in \cite{chronis-catania}, demonstrate the RCPD's behavior in a real test-bed deployment. 
\par Up to now there are only a handful of proposals addressing the challenges of new flexible networking and cloud architectures accounting for content popularity. Exceptions include \cite{cran} in which a logistic-loss machine learning approach to content popularity prediction is applied for a Fog RAN environment, and, our recent papers \cite{chronis-catania} and \cite{GLOBECOM}. In  \cite{chronis-catania}, the algorithm -- outlined in \cite{GLOBECOM} and presented extensively in the present -- is integrated into an elastic CDN framework based on lightweight cloud capabilities using Unikernels. \cite{chronis-catania} focuses on the platform details rather than on the CP algorithm; it confirms experimentally the suitability of the latter for relevant flexible network and cloud architectures. The first detailed description of the proposed CP detection algorithm is presented in the following Sections, along with a rich set of validation results.
We elaborate on the two phases of the RCPD in Sections III and IV respectively and provide the corresponding pseudo-code.       

\section{Training (Off-line) Phase}
\par In this Section, the training phase of the algorithm is discussed and the fundamental components of the off-line scheme are presented. We note that standard off-line CP schemes can only detect a single CP. To address the issue of detection of multiple CPs, we modify the basic algorithm with a novel time-series segmentation heuristic, that belongs to the family of binary segmentation algorithms. 

\subsection{Basic Off-line Approach} 

\par Let $\lbrace{X_n:n\in\mathbb{N}}\rbrace$ be a sequence of $r$- dimensional random vectors (r.v.). The first dimension represents the number of views for a specific video content within a time period $n\in\{1,\ldots, N\}$, while the other dimensions could be optionally used to represent other content popularity features, such as likes, comments, etc. We assume that $X_1,...,X_N$ can be written as, 
\begin{equation}\label{eq:1}
X_n=\mu_n+Y_n, \quad1\leqslant{n}\leqslant{N}
\end{equation}
where $\lbrace{\mu_n:n\in\mathbb{N}}\rbrace$ is the mean value of video visits, $\lbrace{Y_n:n\in\mathbb{N}}\rbrace$ a random component with zero mean ${E}\left[Y_{n}\right]=0$ and positive definite covariance matrix, ${E}\left[Y_n{Y^T_n}\right]=\Sigma$, while ${E}[\cdot]$ denotes expectation. We further assume that the time-series is $t$-dependent, implying that for $t_1, t_2, t \in \mathbb{N} $, $Y_{t_1}$ is independent of $Y_{t_2}$ if $\left|t_1-t_2\right|>t.$   

The model in (\ref{eq:1}) and the underlying assumption of $t-$dependence are in agreement with statistical characterizations of the distribution of visits, which have been shown in numerous analyses to follow either a Zipf \cite{phd-31} or a Zipf-Mandelbrot \cite{phd-38} distribution for both commercial and user-generated content. Furthermore, it is confirmed in the real YouTube datasets used in the present work through the evaluation of the time-series's Hurst exponents, as will be discussed in Section \ref{subsec:Real Data}. 

\par The off-line analysis tests the constancy (or not) of the mean values up to the current time $N$. Hence, we define the following null hypothesis of constant mean, 
    	$$H_0:\quad\mu_{1}=\ldots=\mu_{N},$$
against the alternative,
$$H_1: \quad\mu_{1}=\ldots=\mu_{k_{off}^{*}}\neq\mu_{k_{off}^{*}+1}=\ldots=\mu_N,$$
indicating that the mean value changed at the unknown (time) point $k_{off}^{*}\in\lbrace 1,\ldots ,N\rbrace$.
\par Considering (\ref{eq:1}) and the corresponding assumptions for the stochastic process $X_n$, we develop a non-parametric CUSUM test statistic following \cite{b1}. The test statistic $TS_{off}$, can be viewed as a max-type procedure,
\begin{equation}\label{eq:2}
TS_{off}=\max\limits_{1\leqslant{n}\leqslant{N}}{{C^T_n}\widehat{\Omega}_N^{-1}{C_n}},
\end{equation}
where the parameter $C_n$ is the retrospective CUSUM detector,
\begin{equation}\label{eq:3}
C_n=\dfrac{1}{\sqrt{N}}\left(\sum_{i=1}^{n}{X_i}-n\overline{X}_{1,N}\right),
\end{equation}
while $                     \overline{X}_{1,N}=\frac{1}{N}\sum_{i=1}^{N}{X_i}$ denotes the sample mean. $\widehat{\Omega}_N$ represents a suitable estimator of the long-run covariance $\Omega$, where
\begin{equation}\label{eq:4}
{\Omega}=\sum_{i=-\infty}^{\infty}{\mathbf{Cov}\left(X_nX_{n-i}\right)}.
\end{equation}
The estimator should satisfy,
\begin{equation}\label{eq:5}
\widehat{{\Omega}}_N{\xrightarrow{P}}\Omega
\end{equation}
where ${\xrightarrow{P}}$ denotes convergence in probability. 
\par Several estimators have been proposed in the literature that satisfy (\ref{eq:5}), including kernel-based \cite{b2}, bootstrap-based \cite{b3}, etc. Considering our requirement for real-time detection (low computational time), a kernel-based estimator is more suitable; in this context, we employ the Bartlett estimator, so that
\begin{equation}\label{eq:6}
\widehat{\Omega}_N=\widehat{\Sigma}_0+\sum_{w=1}^{W}k_{BT}{\left(\frac{w}{W+1}\right)}\left(\widehat{\Sigma}_w+\widehat{\Sigma}^T_w\right),
\end{equation}
which satisfies (\ref{eq:5}), while 
the function $k_{BT}(.)$ corresponds to the Bartlett weight,
\begin{equation}\label{eq:7}
k_{BT}\left(x\right)= \begin{cases} 
         1-\abs{x}, \hspace{3mm} \text{for } \abs{x}\leqslant1 \\
0, \hspace{12mm} \text{otherwise} 
   \end{cases},
\end{equation}
and $\widehat{\Sigma}_w$ denotes the empirical auto-covariance matrix for lag $w$, 
\begin{equation}\label{eq:8}
\widehat{\Sigma}_w=\frac{1}{N}\sum_{n=w+1}^{N}\left(X_n-\overline{X}\right)\left(X_{n-w}-\overline{X}\right)^T.
\end{equation}
Finally, we chose $W=\log_{10}(N)$ as in \cite{b2}.
\par The long-run covariance is involved in the test statistic to incorporate the dependence structure of the r.v. into the statistical analysis, through the integration of second order statistical properties. This approach is suitable for the targeted context since we avoid a restrictive assumption for the dependence structure of the observations. 
\par Going back to the basic question of rejecting or not $H_0$, we need to obtain critical values, denoted by $cv_{off}$, for the test statistic. We approach this issue by considering the asymptotic distribution of the test statistic under $H_0$,
\begin{equation}\label{eq:9}
TS_{off}{\xrightarrow{D}}cv_{off}=\sup_{0\leqslant{t}\leqslant{1}}{\sum_{j=1}^{r}}{B_j^2(t)}\quad{(N\rightarrow\infty)},
\end{equation}
where $\xrightarrow{D}$ denotes convergence in distribution, $\left({B_j(t):t\in\left[0,1\right]}\right), \textbf{ } 1\leqslant{j}\leqslant{r},$ are independent standard Brownian bridges $B(t)=W(t)-tW(1)$, and $W(t)$ denotes the standard Brownian motion with mean $0$ and variance $t$. The critical values for several significance levels $\alpha$ can be computed using Monte Carlo simulations that  approximate the paths of the Brownian bridge on a fine grid. The last step is to estimate the unknown CP, defined previously as $k_{off}^*$, under $H_1$, given by:
\begin{equation}\label{eq:10}
\hat{k}^*_{off}=\dfrac{1}{N}{\argmax_{1\leqslant{n}\leqslant{N}}{TS_{off}}}.
\end{equation}

\subsection{Extended Off-line Approach}   
The above hypothesis test identifies the existence of at most one CP and does not ensure that the sample remains statistically stationary in either direction of the detection. In particular, by construction (see (\ref{eq:2})),  the off-line test statistic detects the CP with the highest magnitude. Therefore, for the detection of multiple CPs we need to rephrase the hypothesis test $H_1$, as follows:
\begin{equation}
\begin{aligned}
\nonumber
H_1: \quad\mu_{1}=\ldots=\mu_{k_1}\neq\mu_{k_1+1}=\ldots=\mu_{k_2}\neq\ldots\\
\cdots\neq\mu_{k_{\tau-1}+1}=\ldots=\mu_{k_\tau}\neq\mu_{k_{\tau}+1}=\ldots=\mu_N.
\end{aligned}
\end{equation}
\par A greedy technique to identify multiple CPs is the {binary segmentation} (BS) algorithm. The standard BS algorithm relies on the general concept of binary segmentation and is an extension of the single CP estimator. First, a single CP is searched for in the time-series. In case of no change, the procedure stops and $H_0$ is accepted. Otherwise, the detected CP is used to divide the time-series into two segments in which new searches are performed. The procedure is iterated until no more CPs are detected. The BS algorithm is lightweight (computational time $O(N{\log}N)$), while its conceptual simplicity leads to efficient implementations. On the other hand, it has been shown in the literature \cite{lav}, \cite{ange},
that the standard BS algorithm tends to overestimate the number of CPs, as it does not cross-validate them after their detection. 
\par In the extended off-line approach, we propose the modification of the standard BS with a cross-validation step of the estimated CPs. The cross-validation step is similar to that used in the {iterative cumulative sum of squares} (ICSS) segmentation algorithm \cite{b8}, which is used to search for CPs on the marginal variance of independent and identically distributed (i.i.d.) r.vs. In the extended off-line algorithm we consider the CPs estimated from the standard BS in pairs and check if $H_0$ is rejected in the segment delimited by each pair. If $H_0$ is not rejected in a particular segment, then no change can be detected in it; as a result, all CPs that fall in the respective segment are eliminated. 
The improvement, in terms of accuracy, is shown through simulation results in Section IV. The pseudo-code of the modified BS algorithm is given in \textit{Algorithm 1}; note that we integrate the algorithm with the test statistic $TS_{off}$, given in equation (\ref{eq:2}) and the corresponding critical value ($cv_{off}$) given in (\ref{eq:9}). 
    

\newcommand{\alg}{\texttt{algorithmicx}}
\newcommand{\old}{\texttt{algorithmic}}
\newcommand{\BinSeg}{Modified Binary Segmentation (MBS)}
\newcommand{\BS}{MBS}
\newcommand\ASTART{\bigskip\noindent\begin{minipage}[b]{0.5\linewidth}}
\newcommand\ACONTINUE{\end{minipage}\begin{minipage}[b]{0.5\linewidth}}
\newcommand\AENDSKIP{\end{minipage}\bigskip}
\newcommand\AEND{\end{minipage}}
\alglanguage{pseudocode}
\begin{algorithm}[t]
\caption{\BinSeg}\label{BS}
\begin{algorithmic}[1]
\Procedure{\BS}{start,end,A} 
\State $;$ A: BS method selection (0: standard, 1: modified) 
\State $;$ $TS_{off}$: the off-line test statistic (eq. \ref{eq:2})
\State $;$ ${cv}_{off}$: the critical value (eq. \ref{eq:9})
\State $;$ $\hat{k}^{*}_{off}$: the identified CP (eq. \ref{eq:10}) 
\State calculate \texttt{$TS_{{off}}(start,end)$} and \texttt{$cv_{off}$}
\If{$TS_{off}(start,end)>cv_{off}$}
\State{calculate ${\hat{k}^{*}_{off}}$ and store it in array $s$}
\State{MBS(start,${\hat{k}^{*}_{off}}$,0)}
\State{MBS(${\hat{k}^{*}_{off}}$+1,end,0)}
\EndIf
\If{$\texttt{array\_length}(s)>{0} \text{ and A=1}$}
\State $\widehat{S}\gets$ ${{\lbrace{1}\rbrace}\cup{\lbrace{{s}}\rbrace}\cup{\lbrace{N}\rbrace}}$ $;$ N: the time-series length
\For {i=2:N-1}
\State{MBS($\widehat{S}_{i-1},\widehat{S}_{i+1}$,0)}
\State{keep in $l$ the validated CPs only}
\EndFor
\EndIf
\EndProcedure
\end{algorithmic}
\end{algorithm}




\section{On-line Phase}
\par In this Section, we describe the on-line scheme that includes: (i) two alternative CUSUM-type approaches for the detection of a change in the mean; and (ii) two alternative approaches to estimate the direction of a change.  
\subsection{On-line Analysis}
	\par We rewrite equation (1) in the form,
\begin{equation}\label{eq:11}
X_n= \begin{cases} 
         \mu+Y_n, \hspace{11mm} n=1,\ldots,m+k^*-1 \\
\mu+Y_n+I, \hspace{5mm} n=m+k^*,\ldots 
   \end{cases}
\end{equation}
where $\mu$, $I\in\mathbb{R}^{r}$ represents the mean parameters before and after the unknown time of possible change $k^*\in\mathbb{N^*}$ respectively. As a reminder, the first dimension of the time-series represents the video views; the rest could be likes, comments, etc., and $\lbrace{Y_n: n\in\mathbb{N}}\rbrace$ is a random component. The term $m\in\mathbb{N}$ denotes the length of the training period, i.e., an interval of length $m$ over the historical period during which the mean is assumed to remain unchanged, so that,
\begin{equation}\label{eq:12}
\mu_1=\dots=\mu_m.
\end{equation}
To satisfy this assumption, the modified off-line CP test previously presented is run in order to identify a suitable $m$. With $m$ determined, the on-line procedure can be used to check whether (\ref{eq:12}) holds as new data become available. 

In the form of a statistical hypothesis test, the on-line problem becomes,
\begin{equation}\label{eq:13}
\begin{aligned}
H_0: I=0,\\
H_1: I\neq0.
\end{aligned}
\end{equation}
\par The on-line sequential analysis belongs to the category of stopping time stochastic processes. In general, a chosen on-line test statistic ${TS_{on}(m,l)}$ and a given threshold $F(m,l)$ define the stopping time $\tau(m)$: 
\begin{equation}\label{eq:14}
\tau{\left(m\right)}= \begin{cases} 
         \min\lbrace{l\in{\mathbb{N}}: {TS_{on}(m,l)}{\geqslant{F(m,l)}}}\rbrace,\\
\infty, \text{ if } {TS_{on}(m,l)}{<{F(m,l)}} \text{ }\forall{l\in{\mathbb{N}}},
   \end{cases}
\end{equation}
implying that $TS_{on}(m,l)$ is calculated  on-line  for  every
$l$ in  the  monitoring  period. The procedure stops if the test statistic exceeds the value of the threshold function $F(m,l)$. As soon as this happens, the null hypothesis is rejected and a CP is detected.
The following properties should hold for $\tau(m)$, 
\[ \lim_{m\to\infty} Pr{\left\lbrace\tau(m)<\infty|H_0\right\rbrace=\alpha},\]
ensuring that the probability of false alarm is asymptotically bounded by $\alpha\in\left(0,1\right)$, and,
\[ \lim_{m\to\infty} Pr{\left\lbrace{\tau(m)<\infty|H_1}\right\rbrace=1},\]
ensuring that under $H_1$ the asymptotic power of the statistical test is unity. 
The threshold $F(m,l)$ is given by,
\begin{equation}\label{eq:15}
F(m,l)={cv_{on,a}}{g}{\left({m,l}\right)},
\end{equation}
where: (i) the critical value $cv_{on,a}$ is determined from the asymptotic behavior of the stopping time procedure under $H_0$ by letting $m\rightarrow\infty$; and (ii) the weight function,
\begin{equation}\label{eq:16}
g(m,l)=\sqrt{m}\left(1+\frac{l}{m}\right)\left(\frac{l}{l+m}\right)^\gamma
\end{equation}
depends on the sensitivity parameter $\gamma\in\left[0,1/2\right)$.    
\par We use two different CUSUM approaches; the standard \cite{b10}, with test statistic denoted by  $TS^{ct}_{on}$, and, the ratio-type \cite{b11}, with test statistic denoted by $TS^{rt}_{on}$.  Their corresponding critical values are denoted by $cv^{ct}_{on,a}$ and $cv^{rt}_{on,a}$, respectively, and their stopping rules by $\tau_{ct}(m)$ and $\tau_{rt}(m)$, correspondingly. Both tests are based on the sequential CUSUM detector, $E(m,l)$,
    \begin{equation}{\label{onlC}}
    E(m,l)=\left(\overline{X}_{m+1,m+l}-\overline{X}_{1,m}\right)
    \end{equation}

The standard CUSUM test is expressed as:
\begin{equation}\label{eq:17}
TS^{ct}_{on}(m,l)={l\widehat{\Omega}^{-\frac{1}{2}}_{m}}{E(m,l)},
\end{equation}
where $\widehat{\Omega}_{m}$ is the estimated long-run covariance, defined as in (4), that captures the dependence between observations. Then, the stopping rule $\tau_{ct}(m)$, is defined as: 
\begin{equation}\label{s.t1}
\tau_{ct}(m)=\min\lbrace{l\in\mathbb{N}:\Vert{TS^{ct}_{on}(m,l)}\Vert_{1}\geq{cv^{ct}_{on,a}}g(m,l)\rbrace},
\end{equation}
where the $\ell_{1}$ norm is involved to modify $TS^{ct}_{on}$ so that it can be compared to a one dimensional threshold function.   
The critical value, $cv_{on,a}^{ct}$, is derived from the asymptotic behavior of the stopping rule under $H_{0}$:
\begin{equation}\label{eq:18}
\begin{split}
\lim_{m\to\infty} Pr {\lbrace\tau(m)<\infty\rbrace}=\\ 
& \!\!\!\!\!\!\!\!\!\!\!\!\!\!\!\!\!\!\!\!\!\!\!\!\!\!\!\!\! =\lim_{m\to\infty} Pr {\left\lbrace{\sup_{1\leqslant{l}\leqslant{\infty}}}{\frac{\Vert{TS^{ct}_{on}(m,l)}\Vert_{1}}{g(m,l)}}{>cv^{ct}_{on,\alpha}}\right\rbrace} \\
& \!\!\!\!\!\!\!\!\!\!\!\!\!\!\!\!\!\!\!\!\!\!\!\!\!\!\!\!\! =Pr \left\lbrace\sup_{t\in\left[0,1\right]}{\frac{\Vert{W(t)}\Vert_{1}}{t^\gamma}}>cv^{ct}_{on,\alpha}\right\rbrace=\alpha.
\end{split}
\end{equation}
\par Unlike standard CUSUM tests, ratio type statistics do not require to estimate the long-run covariance and are also considered for this reason in this analysis. The precise form of the chosen statistic is given in the following quadratic form,
\begin{equation}\label{eq:19}
\begin{split}
TS^{rt}_{on}(m,l)=\frac{l^2}{m}{E^T(m,l)}\\
& \!\!\!\!\!\!\!\!\!\!\!\!\!\!\!\!\!\!\!\!\!\!\!\!\!\!\!\!\!\!\!\!\!\!\!\!\!\!\!\!\!\!\!\!\!\!\!\!\!\!\!\!\!\!\!\!\!\!\!\!\!\!\!\!\!\!\!\!\! {\left\lbrace{\frac{1}{m^2}\sum_{j=1}^{m}{j^2}\left(\overline{X}_{1,j}-\overline{X}_{1,m}\right)\left(\overline{X}_{1,j}-\overline{X}_{1,m}\right)^{T}}\right\rbrace}^{-1}{E(m,l)},
\end{split}
\end{equation}
with its equivalent stopping rule, 
\begin{equation}
\tau_{rt}(m)=\min\lbrace{l\in\mathbb{N}:TS^{rt}_{on}\geq{cv^{rt}_{on,a}g^{2}(m,l)}\rbrace}.
\end{equation}
Similarly to the standard CUSUM, the critical value, $cv_{on,a}^{rt}$, is estimated by,
\begin{equation}\label{eq:20}
\lim_{m\to\infty} Pr {\lbrace\tau(m)<\infty\rbrace} 
 =Pr \left\lbrace\sup_{t\in\left[0,\infty\right)}{\Delta_{\gamma}(t)}>cv^{rt}_{on,\alpha}\right\rbrace=\alpha,
\end{equation}
where, 
\newline
$ 
\displaystyle{{\Delta_{\gamma}(t)}=\frac{1}{\eta_{\gamma}^{2}(t)}{B^T(1+t)}{\left(\int_{0}^{1}B(r)B^T(r)dr\right)}^{-1}B(1+t),}
$ 
$
\displaystyle{{\eta_{\gamma}^{2}(t)}={\left(1+t\right)\left(\frac{t}{1+t}\right)}^{\gamma}} ,
$
\newline
and $B(t)$ is a standard Brownian bridge, $t\in\left[0,\infty\right).$
\par Similarly to the off-line case, the on-line critical values for both test statistics can be computed using Monte Carlo simulations, considering that,      
    \begin{equation}\label{cv_standard}
    \begin{split}
    cv^{ct}_{on,\alpha}=\sup_{t\in\left[0,1\right]}{\frac{W(t)}{t^\gamma}},
    \end{split}
    \end{equation}
    \begin{equation}\label{cv_ratio}
    \begin{split}
    cv^{rt}_{on,\alpha}=\sup_{t\in\left[0,\infty\right)}{\Delta_{\gamma}(t)}.
    \end{split}
    \end{equation}
    The estimated on-line CP, $\hat{k}^*_{on}$, is derived directly from the value of the stopping time $\tau(m)$, as,
    \begin{equation}\label{on_line cp}
    \hat{k}^{*}_{on}=m+\lbrace\tau(m)|\tau(m)<\infty\rbrace  .
    \end{equation}
    
\subsection{Trend Indicator}


\par Considering the on-line procedure, the hypothesis $H_{1}$ in (13) is two-tailed because the test statistics $TS^{rt}_{on}$ and $TS^{ct}_{on}$ are formulated in a quadratic form and a $\ell_{1}$ norm, respectively. This means that the stopping time rule $\tau_{ct}(m)$ (or $\tau_{rt}(m)$) cannot be an indicator of the direction of a detected change. Thus, to estimate the direction of a change we introduce two indicators: i) based on the CUSUM detector in (\ref{onlC}), denoted by $TI_{ts}$; and ii) based on the moving average convergence divergence (MACD) filter \cite{b5}, denoted by $TI_{f}$.   
\par Focusing on $TI_{ts}$, the indicator is directly derived from the form of the sequential CUSUM detector $E(m,l)$. As shown in (17), the detector compares the mean value of the observations that are collected on-line for a chosen monitoring period $l$, with the mean value of a subsample of the historical data over the predetermined training sample. Hence, for a detected CP, we have that, 
\begin{equation}
\begin{cases}
E(m,l)>{0}, \text{ denotes an upward change}\\
E(m,l)<{0}, \text { denotes a downward change}
\end{cases}.
\end{equation}
\par However, in certain cases, limiting the window over which the direction of a change is estimated to the immediate neighbourhood of a detected CP can be unreliable due to the continuous variability of the time-series. In such cases, we have to estimate the direction of a change by incorporating more elaborate filters; in this context, we estimate the direction of detected changes by applying the MACD indicator. The MACD is based on an exponential moving average (EMA) filter, of the form,
\begin{equation}\label{eq:21}
EMA_{p}(n)=\dfrac{2}{p+1}X_{n}+\dfrac{p-1}{p+1}EMA_{p}(n-1),
\end{equation}
with $p$ denoting the lag parameter. The MACD series can be derived from the subtraction from a short $p_2$ lag EMA (sensitive filter) of a longer $p_3$ lag EMA (blunt filter), as described below:
\begin{equation}\label{eq:22}
MACD(n)=EMA_{p_\text{2}}-EMA_{p_\text{3}}.
\end{equation}
The trend indicator $TI_{f}$ is then obtained by the subtraction of a short $p_1$ lag EMA filter of a MACD series from the raw MACD series, as described below
\begin{equation}
TI_{f}(n)=MACD(n)-EMA_{p_\text{1}}(MACD(n)), \label{eq:TI}
\text{ }p_1<p_2<p_3.
\end{equation}
\par In the evaluation of $TI_{f}$ three exponential filters are involved. In essence, $TI_{f}$ is an estimation of the second derivative over an interval around the change (considering that the subtraction of a filtered variable from the variable generates an estimate of its time derivative). In contrast to other works \cite{b5}, we only adopt $TI_{f}$ to characterize the direction from the specific value of $TI_{f}$ at the estimated time of change.  
We announce an upward change if $TI_{f}(\hat{k}^{*}_{on})>0$, otherwise, if $TI_{f}(\hat{k}^{*}_{on})<0$, a downward change. 

Finally, we propose a modification of the trend indicator $TI_f$, converting it from a point estimator to an interval estimator; instead of evaluating $TI_{f}(\hat{k}^{*}_{on})$, we propose to evaluate the trend indicator 
at a time interval $(\hat{k}^{*}_{on},\hat{k}^{*}_{on}+h)$, where $h$ is a threshold parameter:
\begin{equation}
\begin{split}
TI_{f}(\hat{k}^{*}_{on},h)=\sum_{l=\hat{k}^{*}_{on}}^{\hat{k}^{*}_{on}+h}TI_{f}(l).
\label{MTI}
\end{split}
\end{equation}
The proposed $TI_{f}(\hat{k}^{*}_{on},h)$ modification improves the estimator's accuracy; the calculation of the sum of a multitude of observations, after a CP, can smooth out a potential false one-point estimation, especially in the case of small changes.
	
\newcommand{\OnlP}{The Real-time CP Detector (RCPD)}
\newcommand{\OLA}{RCPD}
\alglanguage{pseudocode}
\begin{algorithm}[t]
\caption{\OnlP}\label{OLA}
\begin{algorithmic}[1]
\Procedure{\OLA}{$X_n$,$m_s$,k} 
\State $;$ $X_n$: time-series of video views
\State $;$ $m_s$: running end of training period
\State $;$ $m$: training period
\State $;$ $l$: monitoring time frame
\State $;$ $d$: period assuming no change
\State $;$ $TS_{on}$: on-line test statistic (eq. \ref{eq:17} or \ref{eq:19})
\State $;$ $cv_{on}$: critical value (eq. \ref{cv_standard} or \ref{cv_ratio})
\State $;$ $\hat{k}^{*}_{on}$: the estimated on-line CP (eq. \ref{on_line cp})
\State $;$ TI: trend indicator ($TI_{ts}$ or $TI_f$)
\For{n in $X_n$}
\If{$n=m_s$}
\State $s$=MBS(1,$m_s$,1) $;$ calculate off-line CPs
\If{$\texttt{array\_length}(s)>0$}
\State m=$\lbrace{\max(s),m_s}\rbrace$ $;$ $max(s)$ is the latest CP
\Else
\State m=$\lbrace{max(1,m_s-u),m_s}\rbrace$ $;$ $u$ a large value  
\EndIf
\ElsIf{$m_s<n<m_{s}+l$}
\State calculate \texttt{{TS$_{on}$(m,l)}}
\If{\texttt{TS$_{on}$(m,l)}$>$cv$_{on}$}
\State calculate {\texttt{TI}}
\State signal CP and estimated direction
\State $m_s=\widehat{cp}_{on}+d$ $;$ keep a distance from $\widehat{cp}_{on}$
\EndIf
\ElsIf{$n=m_{s}+l$} 
\State $m_s=m_s+l$ $;$ start a new training period
\EndIf
\EndFor
\EndProcedure
\end{algorithmic}
\end{algorithm}

\subsection{Overall Algorithm}
\par We outline in \textit{Algorithm 2} the RCPD algorithm, as a combination of the off-line and the on-line phase, in the form of pseudo-code. Beginning from the initial value set for the monitoring starting period, denoted by $m_s$, the modified off-line algorithm is applied over the whole historical period; the training period $m$ is then defined as the interval elapsed from the last detected off-line CP (if one exists) to $m_s$. In pseudo-code this step is described in lines $14-18$. As a second step, the on-line test statistic, $TS_{on}(m,l)$ in (14), is applied 
for a specified monitoring time frame $l$. If a content popularity change is detected at time instance $\hat{k}^{*}_{on}$, the trend indicator subroutine is called to reveal the direction of change.\footnote{In the load balancing scenario discussed in Section VII, in the case of an increase in the content popularity a new content cache is being deployed, while conversely a decrease leads to the removal of an existing cache.} At this point the procedure stops and a new starting point for the monitoring window is defined as $m_s=\hat{k}^{*}_{on}+d$, where $d$ is a constant value specifying a period assuming no change. This step is described in lines $19-29$. Otherwise, if no change is detected after a maximum of $l$ instances, the procedure restarts from the last time point, $m_s=m_s+l$. 
    
\section{Validation of the RCPD Using Synthetic Data}\label{results}
	
\par In this Section, we validate the performance of the overall algorithm by performing a series of four different experiments on synthetic data. The use of synthetic data allows us to regulate the parameters of the time-series in terms of mean changes and thus obtain quantitative metrics for the performance of the proposed algorithms. 
  
\par The choice of the time-series model for the generation of the synthetic data is based on the fact that several studies have shown that ARMA models capture very well content popularity evolution. For example, in \cite{survey-32} it has been concluded that an ARMA model can efficiently describe the daily access patterns of YouTube content, based on an extensive analysis of $100,000$ videos. Similarly, in \cite{Hassine} an ARMA model has been proposed for the estimation of the popularity of video content. Motivated by these findings, for the validation of the proposed algorithm we use an ARMA$(1,1)$ time-series.
We generate $1,000$ time-series of length $N=600$ samples. Without loss of generality, we assume an initial mean value $\mu_0=0$, noting that the performance of the RCPD is independent of the initial mean value and only depends on the magnitude of the variation of the mean value before and after a CP.
 
\begin{table}[t]
\label{ta:BS}
\begin{center}
\renewcommand{\arraystretch}{1.4}
\setlength{\tabcolsep}{8pt}
\centering
\caption{Percentage of the successful CP detections for the standard and modified BS algorithm}
\begin{tabular}{|l|cc|cc}
\hline
 & \multicolumn{2}{|c|}{Test 1: two CPs} & \multicolumn{2}{|c|}{Test 2: four CPs} \\
 \hline
 \multicolumn{1}{|l|}{$\mu$} & \multicolumn{1}{c}{BS} & \multicolumn{1}{c|}{modified BS} & \multicolumn{1}{c}{BS} & \multicolumn{1}{c|}{modified BS}\\
\hline
 & \multicolumn{2}{c|}{True (false) alarm rate} & \multicolumn{2}{c|}{True (false) alarm rate} \\
 \hline
\multicolumn{1}{|l|}{$\mu_{1}$=1} & \multicolumn{1}{c}{0.94\textit{ }(0.06)} &\multicolumn{1}{c|}{0.95\textit{ }(0.05)} & \multicolumn{1}{c}{0.5\textit{ }(0.258)} & \multicolumn{1}{c|}{0.7\textit{ }(0.05)}\\

\multicolumn{1}{|l|}{$\mu_{2}$=1.5} & \multicolumn{1}{c}{0.95\textit{ }(0.05)} &\multicolumn{1}{c|}{0.95\textit{ }(0.05)} & \multicolumn{1}{c}{0.5\textit{ }(0.258)} & \multicolumn{1}{c|}{0.9\textit{ }(0.08)}\\

\multicolumn{1}{|l|}{$\mu_{3}$=2} & \multicolumn{1}{c}{0.95\textit{ }(0.05)} &\multicolumn{1}{c|}{0.95\textit{ }(0.05)} & \multicolumn{1}{c}{0.47\textit{ }(0.53)} & \multicolumn{1}{c|}{0.9\textit{ }(0.1)}\\
\hline
\end{tabular}    
\end{center}      
\end{table}    
 
In the first experiment, we begin with a comparison of the standard BS to the proposed modified BS algorithms described in Section II-B. We perform two tests; in the first test we introduce two CPs at the instances $k^{*}_{i}=(iN)/3, \textit{ } i=1,2$, while in second test, we introduce four CPs at $k^{*}_{i}=(iN)/5, \textit{ } i=1, \ldots,4$. 
The two tests are repeated for three different values of the magnitude of a change $\mu_{1}=1, \textit{ }\mu_2=1.5, \mu_3=2 $, i.e., we randomly increase or decrease the mean value by $\mu_j, \textit{ }j=1,\ldots,3$ at the time of change. Table I summarizes our findings regarding the true and false alarm rates of the two algorithms. 

Both the standard and the modified BS algorithms provide similar true alarm rates, exceeding $94\%$, in the first test. On the contrary, in the more challenging second test, the superiority of the modified BS over the standard BS algorithm is clear. The modified BS algorithm achieves true alarm rates in excess of $70\%$, even in the demanding scenario of a relatively small change in the mean $\mu_1=1$. On the other hand, the standard BS algorithm has in all cases a true alarm rate of less than $50\%$, rendering any CP detection highly questionable. The second test confirms that the standard BS algorithm is prone to an overestimation of the number of CPs as shown by the high false alarm rates (in excess of $25\%$ in all cases), an issue that can be effectively addressed by the modified BS algorithm which scores false alarm rates below $10\%$. 

\begin{table}[t]
\begin{center}
\renewcommand{\arraystretch}{1.4}
\setlength{\tabcolsep}{10pt}
\centering
\caption{Success rates of trend indicators}
\begin{tabular}{|l|cc|cc}
\hline
 & \multicolumn{2}{|c|}{Test 1: two CPs} & \multicolumn{2}{|c|}{Test 2: four CPs} \\
 \hline
 \multicolumn{1}{|l|}{$\mu$} & \multicolumn{1}{c}{$TI_{ts}$} & \multicolumn{1}{c|}{$TI_{f}$} & \multicolumn{1}{c}{$TI_{ts}$} & \multicolumn{1}{c|}{$TI_{f}$}\\
\hline
 & \multicolumn{2}{c|}{Success rate} & \multicolumn{2}{c|}{Success rate} \\
 \hline
\multicolumn{1}{|l|}{$\mu_{1}$=1} & \multicolumn{1}{c}{0.99} &\multicolumn{1}{c|}{0.99} & \multicolumn{1}{c}{0.99} & \multicolumn{1}{c|}{0.99}\\
\multicolumn{1}{|l|}{$\mu_{2}$=1.5} & \multicolumn{1}{c}{1} &\multicolumn{1}{c|}{1} & \multicolumn{1}{c}{1} & \multicolumn{1}{c|}{1}\\
\multicolumn{1}{|l|}{$\mu_{3}$=2} & \multicolumn{1}{c}{1} &\multicolumn{1}{c|}{1} & \multicolumn{1}{c}{1} & \multicolumn{1}{c|}{1}\\
\hline
\end{tabular}
\end{center}
\end{table}
\begin{table*}[t]
\begin{center}
\renewcommand{\arraystretch}{1.4}
\setlength{\tabcolsep}{13pt}
\centering
\caption{Results of the RCPDs' algorithm CPs detection for one change in the mean value.}
\label{tab:cp_mean_one_1}
\begin{tabular}{|ll|ccccccccccc|}
\hline
\cline{1-13}
 & & \multicolumn{11}{c|}{ARMA(1,1)} \\
\hline
\cline{1-13}
$\mu$ & $l$ & \multicolumn{5}{c}{standard CUSUM} &  & \multicolumn{5}{c|}{ratio-type CUSUM} \\
\cline{1-2}
\cline{3-7}
\cline{9-13}
 &  &  \multicolumn{3}{c}{Number of detected CPs} &  & $\hat{k}^{*}$ &  & \multicolumn{3}{c}{Number of detected CPs} &  & $\hat{k}^{*}$ \\
\cline{3-5}
\cline{9-11}
\cline{7-7}
\cline{13-13}
 &  & 0 & 1 & $>{1}$ &  & med &  & 0 & 1 & $>{1}$ & & med\\
\hline
 & $25$ & 0.95 & 0.05 & 0 & & - &  & 0.95 & 0.05 & 0 & &\multicolumn{1}{c|}{- }\\
$\mu=0$ & $50$ & 0.95 & 0.05 & 0 & & - &  & 0.95 & 0.05 & 0 & &\multicolumn{1}{c|}{- }\\
 & $100$ & 0.94 & 0.06 & 0 & & - &  & 0.95 & 0.05 & 0 & &\multicolumn{1}{c|}{- }\\
 \cline{1-13}
 & $25$ & 0.7 & 0.29 & 0.01 &  & - &  & 0.8 & 0.19 & 0.01 &  & -\\ 
$\mu=0.5$ & $50$ & 0.16 & 0.8 & 0.04 & & 343 & & 0.55 & 0.43 & 0.02 &  & -\\ 
  & $100$ & 0 & 0.93 & 0.07 & & 341 & & 0.2 & 0.76 & 0.04 &  & 348\\ 
\cline{1-13}
 & $25$ & 0.26 & 0.73 & 0.01 & & 332 & & 0.69 & 0.3 & 0.01 &  & -\\
$\mu=0.7$ & $50$ & 0 & 0.96 & 0.04 & & 326 & & 0.3 & 0.65 & 0.05 &  & 328\\
  & $100$ & 0.01 & 0.91 & 0.08 & & 331 & & 0.05 & 0.89 & 0.06 &  & 335\\
\cline{1-13}
 & $25$ & 0.01 & 0.97 & 0.02 & & 327 & & 0.52 & 0.46 & 0.02 &  & -\\
$\mu=1$ & $50$ & 0 & 0.96 & 0.04 & & 316 & & 0.08 & 0.86 & 0.06 &  & 321\\
  & $100$ & 0 & 0.92 & 0.08 & & 321 & & 0 & 0.95 & 0.05 &  & 323\\
\cline{1-13}
 & $25$ & 0.01 & 0.97 & 0.02 & & 323 & & 0.43 & 0.54 & 0.03 &  & 331\\
 $\mu=1.2$ & $50$ & 0 & 0.95 & 0.05 & & 316 & & 0.02 & 0.93 & 0.05 &  & 317\\
  & $100$ & 0 & 0.93 & 0.07 & & 318 & & 0 & 0.93 & 0.07 &  & 318\\
\cline{1-13}
 & $25$ & 0 & 0.97 & 0.03 & & 320 &  & 0.36 & 0.6 & 0.04 &  & 329\\
$\mu=1.5$ & $50$ & 0 & 0.95 & 0.05 & & 310 &  & 0 & 0.94 & 0.06 &  & 313\\
  & $100$ & 0 & 0.93 & 0.07 & & 314 &  & 0 & 0.94 & 0.06 &  & 318\\
\cline{1-13}
 & $25$ & 0 & 0.97 & 0.03 & & 310 & & 0.26 & 0.71 & 0.03 &  & 317\\
$\mu=2$ & $50$ & 0 & 0.95 & 0.05 & & 307 & & 0 & 0.93 & 0.07 &  & 310\\
  & $100$ & 0 & 0.94 & 0.06 & & 310 & & 0 & 0.94 & 0.06 &  & 313\\
\cline{1-13}
\end{tabular}
\end{center}
\end{table*}

\begin{table*}[t]
\begin{center}
\renewcommand{\arraystretch}{1.4}
\setlength{\tabcolsep}{10.5pt}
\centering
\caption{Results of the RCPDs' algorithm CPs detection for two mean changes.}
\label{tab:cp_mean_one_1}
\begin{tabular}{|ll|ccccccccccccc|}
\hline
\cline{1-15}
\multicolumn{2}{|c|}{} & \multicolumn{13}{c|}{ARMA(1,1)}\\ 
\hline
\cline{1-15}
$\mu$ & $l$ & \multicolumn{6}{c}{standard CUSUM} &  & \multicolumn{6}{c|}{ratio-type CUSUM}\\ 
\cline{1-2}
\cline{3-8}
\cline{10-15}
 &  & \multicolumn{3}{c}{Number of detected CPs} &  & $\hat{k}^{*}_1$ & $\hat{k}^{*}_2$ & & \multicolumn{3}{c}{Number of detected CPs} &  & $\hat{k}^{*}_1$ & $\hat{k}^{*}_2$\\  
\cline{3-5}
\cline{7-8}
\cline{10-12}
\cline{14-15}
 &  & $<{2}$ & 2 & $>{2}$ &  & \multicolumn{2}{c}{med} &  & $<{2}$ & 2 & $>{2}$ & & \multicolumn{2}{c|}{med}\\
\hline
 & $25$ & 0.88 & 0.12 & 0 & & - & - &  & 0.95 & 0.05 & 0 & & - & -\\ 
 $\mu_1=0.5$ & $50$ & 0.38 & 0.60 & 0.02 & & 251 & 440 &  & 0.79 & 0.2 & 0.01 &  & - & -\\ 
  & $100$ & 0.1 & 0.87 & 0.03 & & 242 & 443 &  & 0.54 & 0.44 & 0.02 &  & - & -\\  
 \cline{1-15}
  & $25$ & 0.41 & 0.58 & 0.01 & & 230 & 427 &  & 0.9 & 0.1 & 0 &  & - & -\\ 
 $\mu_1=0.7$ & $50$ & 0.06 & 0.91 & 0.03 & & 223 & 427 &  & 0.58 & 0.41 & 0.01 &  & - & -\\ 
  & $100$ & 0.01 & 0.93 & 0.06 & & 227 & 428 &  & 0.25 & 0.72 & 0.03 &  & 231 & 439\\  
 \cline{1-15}
  & $25$ & 0.04 & 0.93 & 0.03 & & 219 & 420 &  & 0.74 & 0.25 & 0.01 &  & - & -\\  
 $\mu_1=1$ & $50$ & 0.03 & 0.93 & 0.04 & & 215 & 419 &  & 0.26 & 0.71 & 0.03 &  & 221 & 423\\  
  & $100$ & 0 & 0.94 & 0.06 & & 217 & 420 &  & 0.05 & 0.9 & 0.05 &  & 220 & 424\\  
 \cline{1-15}
 & $25$ & 0.01 & 0.96 & 0.03 & & 214 & 414 &  & 0.56 & 0.42 & 0.02 &  & - & -\\ 
 $\mu_1=1.2$ & $50$ & 0 & 0.95 & 0.05 & & 212 & 416 &  & 0.17 & 0.79 & 0.04 &  & 215 & 428\\  
  & $100$ & 0 & 0.94 & 0.06 & & 217 & 420 &  & 0.02 & 0.93 & 0.05 &  & 216 & 421\\  
 \cline{1-15}
 & $25$ & 0 & 0.98 & 0.02 & & 211 & 411 &  & 0.33 & 0.63 & 0.04 &  & 213 & 417\\  
 $\mu_1=1.5$ & $50$ & 0 & 0.94 & 0.06 & & 209 & 413 &  & 0.1 & 0.85 & 0.05 &  & 213 & 415\\  
  & $100$ & 0 & 0.94 & 0.06 & & 211 & 415 &  & 0 & 0.96 & 0.04 &  & 216 & 419\\  
 \cline{1-15}
 & $25$ & 0 & 0.98 & 0.02 & & 208 & 407 &  & 0.12 & 0.85 & 0.03 &  & 210 & 412\\  
 $\mu_1=2$ & $50$ & 0 & 0.95 & 0.05 & & 207 & 410 &  & 0.3 & 0.91 & 0.06 &  & 209 & 413\\ 
  & $100$ & 0 & 0.94 & 0.06 & & 209 & 411 &  & 0 & 0.96 & 0.04 &  & 211 & 414\\  
 \hline
 \cline{1-15}
\end{tabular}
\end{center}
\end{table*}    

\par Next, in the second experiment, using the same test sets as above, we measure the success rates achieved by the proposed trend indicators $TI_{ts}$ in (27) and $TI_{f}$ in (31) for $h=0$ (larger thresholds provided the same true identification rates). The results are summarized in Table II. The two trend indicators successfully identify the direction of a change in more than $99\%$ of the cases, which shows that they can be interchangeably employed. In the assessment of the performance using real datasets in Section VI, we solely employ the $TI_{f}$ trend indicator.

\par We proceed by assessing the proposed RCPD algorithm using both the standard and the ratio type CUSUM. In this third experiment, we measure the average number of CPs detected, averaged over $1,000$ simulations when a single CP is introduced in the ARMA time-series at the time instance $\frac{N}{2}=300$.
We consider different values for the magnitude of change $\mu \in \{0, 0.5, 0.7, 1, 1.2, 1.5, 2\}$ and the monitoring window length $l\in \{25, 50, 100\}$. We note that we included the case $\mu=0$ -- which corresponds to the absence of a change -- to evaluate the false alarm rate of the overall algorithm. We omit results with true alarm rates lower than $50\%$ as they are statistically unreliable. In terms of the remaining algorithmic parameters, we have set the minimum distance between two successive CPs to $d=50$,\footnote{This choice is justified by our observations of the minimum distance between successive CPs in real data sets, presented in Section VI.} the sensitivity parameter to $\gamma=0.25$ \cite{g} (we choose a neutral value as the behaviour of $\gamma$ is well studied), and, the significance level to $\alpha=0.05$. In  each test of the third experiment we measure the exact number of CPs detected, tabulated as one the following three values: 
i) $0$ when (falsely\footnote{Except for the $\mu=0$ case.}) no CP is detected; ii) $1$ when (correctly) a single CP is detected; and iii) $> 1$ when (falsely) multiple CPs are detected. Finally, we measure the median of the time instance of the single CP detection, denoted by $\hat{k}^*.$\footnote{We omit the results with true detection rate lower than $50\%$.}
The results of this experiment are presented in Table III and are discussed below. 

Firstly, we observe that both the standard and the ratio type CUSUM achieve very small false alarm rates, inferior to $6\%$ when no CP is inserted, irrespective of the choice of $l$. On the contrary, the choice of $l$ readily affects the algorithm's success rate for $\mu >0$; for small changes in the mean value, $\mu=0.5, 0.7$, a larger monitoring window $l$ increases the algorithm's true alarm rates in identifying correctly the existence of the CP. For medium and high changes in the magnitude of change $\mu=1, 1.2, 1.5, 2$, it is observed that a high true alarm rate -- in excess of $93\%$ for the standard CUSUM -- is achieved, while choosing a smaller $l$ can slightly increase the true alarm rates. As a result, depending on the application, a choice of a larger $l$ can be appropriate if the algorithm is to be employed as a universal CP detector. Alternatively, a smaller $l$ can be chosen when the focus is on the identification of large changes in the mean value, i.e., we are interested primarily in detecting CPs of larger magnitude. 

Secondly, we observe that overall, the ratio type CUSUM is outperformed by the standard CUSUM in all tests. Consequently, the standard CUSUM based detector can be considered as an efficient universal choice. Finally, we observe that the lag between $\hat{k}^{*}$ and the actual instance of change at the point $300$ decreases with increasing $\mu$, ranging from $343$ to $307$, while it appears less sensitive to changes in $l$. This demonstrates that, intuitively, larger magnitude changes can be detected faster. This result is important for load balancing applications as it provides us with the means to quickly respond to significant changes in the network traffic.

Subsequently, in Table IV in the following page, we present the outputs of the fourth experiment in which we assess the performance, averaged over $1,000$ simulations, of the RCPD algorithm when two CPs are inserted in the ARMA time-series. 
We introduce a change at the time instance $k^{*}_1=\frac{N}{3}=200$ and a second CP at the time instance $k^{*}_2=\frac{2N}{3}=400$. We investigate the true and false alarm rates for $\mu \in \{0.5, 0.7, 1, 1.2, 1.5, 2 \}$ and $l\in \{25, 50, 100\}$, while the rest of the parameters retain the values of the third experiment.  In  each test of the fourth experiment we measure the exact number of CPs detected, tabulated as one the following three values: i) $< 2$ when (falsely) less than two CPs are detected, ii) $2$ when (correctly) two CPs are detected, and iii) $> 2$ when (falsely) more than two CPs are detected. Finally, we measure the median of the detection instances of the two CPs, denoted by $\hat{k}^{*}_{1}$ and $\hat{k}^{*}_{2}$, respectively (we omit the results with true detection rate lower than $50\%$). 

Similarly to the third experiment, we observe that increasing $l$ increases the true alarm rates for small magnitudes in the mean changes $\mu=0.5, \textit{ }0.7$, while this trend is reversed in high magnitudes $\mu=1.5, \textit{ }2$. For medium values $\mu=1, \textit{ }1.2$ the effect of $l$ on the true alarm rates is less than $2\%$. Furthermore, in agreement with the outputs of the third experiment, with increasing $\mu$ the algorithms achieve increasingly high success rates, over $93\%$ for the standard CUSUM when $\mu\geq 1$. 

In addition, the superior performance of the standard CUSUM is re-confirmed in all the tests of the fourth experiment. Finally, with respect to the lag in the estimation of the time instances of the CPs, we observe that, as in experiment three, larger magnitude changes can be detected faster, e.g., for $\mu=2$ a lag inferior to $11$ instances is observed for both CPs with the standard CUSUM, irrespective of $l$.

Concluding this Section, we have presented an extensive set of experiments that provide strong evidence for the efficiency of the proposed algorithms. We have explicitly demonstrated the superiority of the modified BS over the standard BS algorithm and confirmed the validity of the proposed trend indicators.  Subsequently, we evaluated the performance of the overall algorithm for various values of $\mu$ and $l$. We have shown that the RCPD algorithm achieves extremely high true alarm rates for larger values of $\mu$, while increasing the length of the monitoring window $l$ can significantly impact the performance for small values of $\mu$. Finally, overall, the standard type CUSUM outperforms the ratio type CUSUM and should be preferred. 


\section{Performance Evaluation Using Real Data}

 In this Section we investigate the performance of the proposed algorithms   using a real dataset provided within the framework of the CONGAS project \cite{congas}; the dataset consists of the number of views of $882$ YouTube videos, observed over $N=1,000$ instances.
 
\subsection{Statistical Properties of the Real Dataset} \label{subsec:Real Data}

\par First, we evaluate the validity of the most important underlying assumption of this analysis, that the content popularity can be modelled as the sum of a constant mean and a weak-dependent ($t$-dependent) stochastic process, as given in (\ref{eq:1}). A first intuitive method to test whether the time-series is short-range dependent (SRD) is through its autocorrelation function (ACF). The ACF for a weakly-stationary process $\{X_{t}:t\in\mathbb{N}\}$ with mean value $\mu$ is given by,
    $$\rho(k)=\frac{(X_{t}-\mu)(X_{t+k}-\mu)}{\sigma^{2}}.$$
Note that if $\sum_{k=-\infty}^{\infty}{\rho(k)}\rightarrow\infty$ the process has long-range dependence (LRD), while if $\sum_{k=-\infty}^{\infty}\abs{\rho(k)}<\infty$ it exhibits SRD. 
To distinguish between these two phenomena, we use the following functional form of the ACF,
$$\rho(k)\sim{C_{i}^{2H-2}}, \textrm{ as}  \textit{ } i\to{\infty},$$ where $C_{i}>0$ and $H\in(0,1)$ is the Hurst exponent characterizing the LRD,i.e., $H\in(1/2, 1)$ indicates the presence of LRD. It is challenging to accurately estimate the Hurst exponent out of real data \cite{clegg} and several methods have been proposed in the literature \cite{bardet}. In this work, we apply two semi-parametric tests, identified as accurate options among others presented in the survey paper \cite{bardet}. The first method uses the discrete second order derivative in the time domain while the second uses the discrete second order derivative in the wavelet domain. Both methods estimate an $H\leq0.5$ for $95\%$ of the YouTube time-series, indicating the validity of our assumptions related to the equation (\ref{eq:1}).

\subsection{Performance of the Off-line Training Phase}
\par First, we test the hypothesis $H_0$ of no change in the mean structure on our dataset. $H_0$ is rejected in approximately $70\%$ of the cases, for a significance level of $a=0.05$. This outcome indicates that CP algorithms can identify changing content dynamics in real times series.
\par Next, we estimate the number of CPs, by applying the extended off-line algorithm. The corresponding results are illustrated in Fig. 1 and indicate a sufficiently high number of content popularity anomalies (i.e., mean changes). Hence, a CP analysis is indeed a suitable tool for content popularity detection.
   
\begin{figure}
\centering
    \begin{subfigure}[b]{0.25\textwidth}       \includegraphics[width=\textwidth]{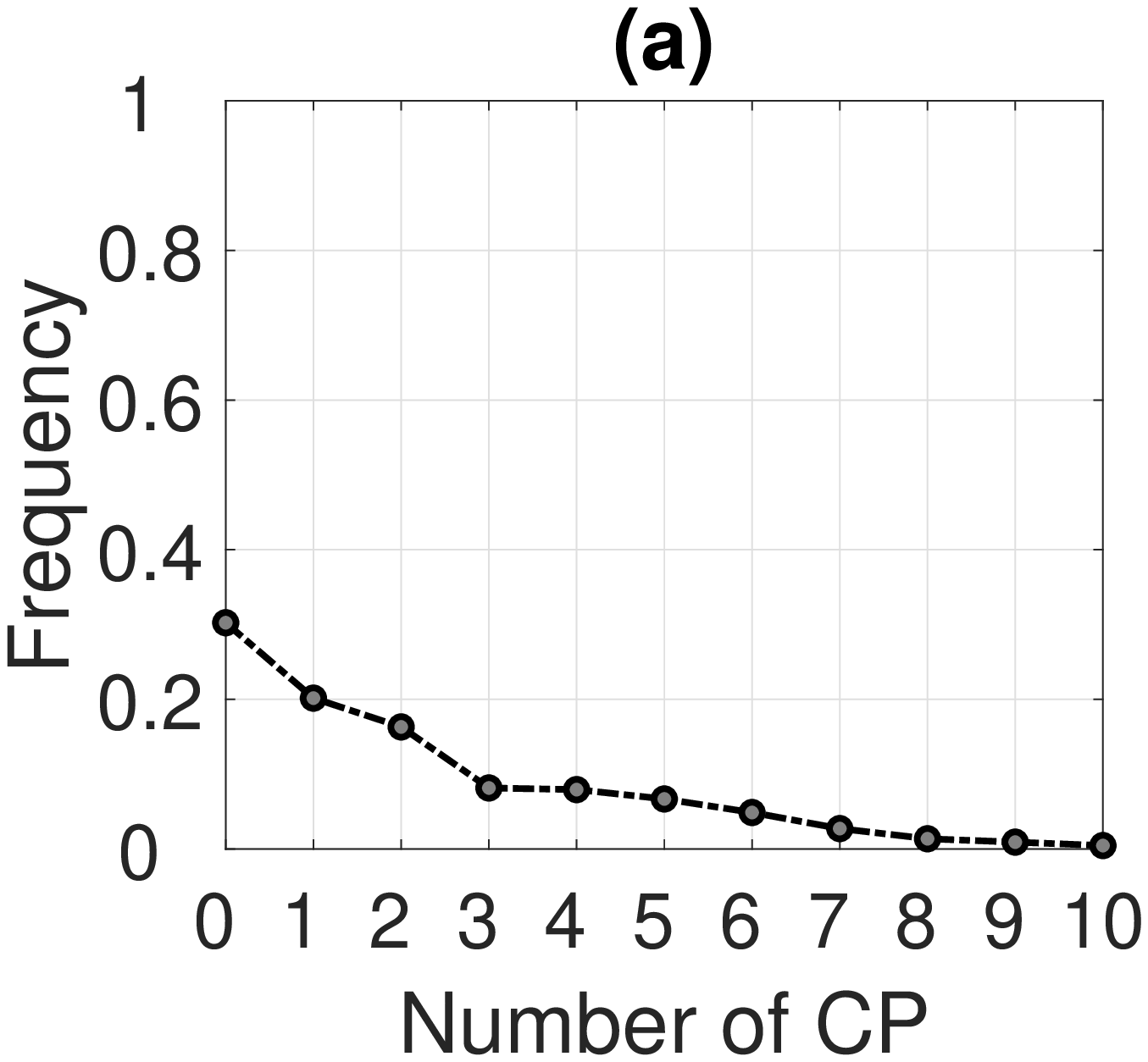}
    \end{subfigure}%
    \begin{subfigure}[b]{0.25\textwidth}
   \includegraphics[width=\textwidth]{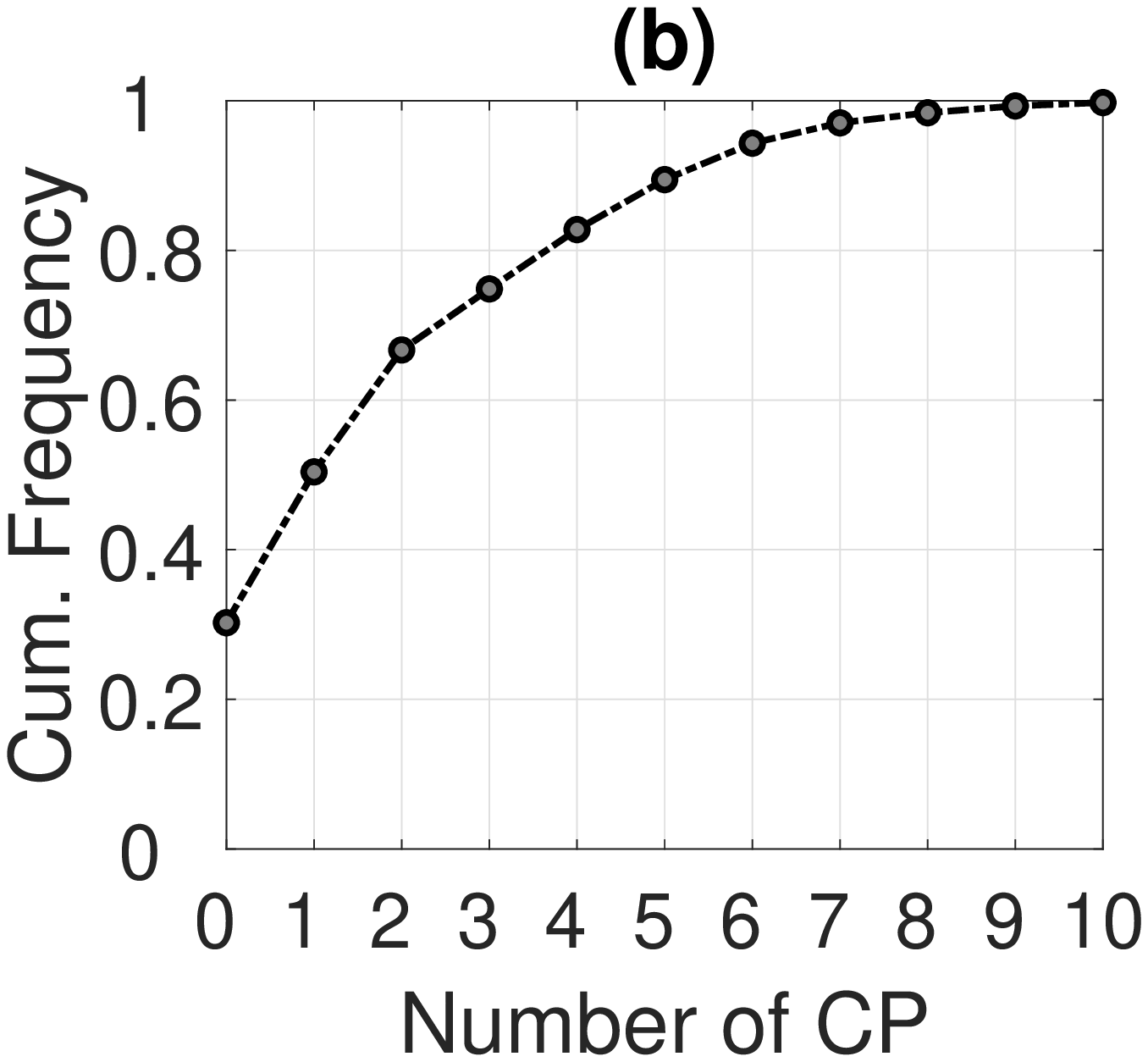}
    \end{subfigure}
    \caption{Estimated a) frequency and b) cumulative frequency of the number of CPs per time-series.} 
\end{figure}

 \begin{table}[t]
\renewcommand{\arraystretch}{1.4}
\setlength{\tabcolsep}{11pt}
\centering
\caption{Success rates of $TI_{f}$ trend indicator}
\label{tab:cp_direction}
\begin{tabular}{cccccc}
\hline
\hline
h & 0 & 3 & 5 & 7 & 10\\
\hline
\multirow{1}{*}{Video Set 1}& 0.69 & 0.91 & 0.95 & 0.97 & 0.98\\
\hline
\multirow{1}{*}{Video Set 2}& 0.90 & 0.99 & 0.99 & 0.99 & 0.99\\
 \hline
\end{tabular}
\end{table}  
\par To evaluate the performance of the proposed trend indicator $TI_{f}$, we need a baseline independent assessment of the direction of change. We declare that a real increase in the mean value of content visit exists if 
\begin{equation}
\mathsf{E}[X(\hat{k}^{*}_{i-1,off}): X(\hat{k}^{*}_{i,off})]<\mathsf{E}[X(\hat{k}^{*}_{i,off}):X(\hat{k}^{*}_{i+1,off})],
\end{equation}
or, that a real decrease in the number of visits exists if 
\begin{equation}
\mathsf{E}[X(\hat{k}^{*}_{i-1,off}):X(\hat{k}^{*}_{i,off})]>\mathsf{E}[X(\hat{k}^{*}_{i,off}):X(\hat{k}^{*}_{i+1,off})],
\end{equation}
where $i=1,\cdots,card(\hat{k}^{*}_{off})$, $\hat{k}^{*}_{0}=1$, $\hat{k}^{*}_{s+1}=N$ and $\mathsf{E}[\cdot]$ denotes the numerical average.
 We test the modified MACD $TI_{f}$ on two sets of videos. The first set, Video Set 1, comprises the whole dataset, while the second set, Video Set 2, comprises only the videos with a considerable average number of visits ($>10$), i.e., for which, $\mathsf{E}[X(1):X(1000)]>10.$

The percentage of successful $TI_{f}$ identifications are tabulated in Table V for five values of the parameter $h$, namely $h=0, \text{ } 3, \text{ } 5, \text{ } 7$ and $10$, where $h$ denotes the $TI_{f}$'s calculation threshold introduced in Section IV-B. Commenting on the results for Video Set 1, the $TI_{f}$ trend indicator works well, except for $h=0$, providing at least $90\%$ correct direction identifications. As expected, as $h$ increases the procedure works better. More specifically, an $h\geq{5}$ parameter choice yields a success rate of $95\%$, while if a more agile estimation is needed then an $h\geq{3}$ still maintains a $91\%$ accuracy. Considering the interim time between consecutive changes, we deduce that an $h\leq{7}$ is preferable. Regarding Video Set 2, we see that the results are highly improved, indicating that the procedure works even better for the most popular videos. In practice, this represents the more interesting scenario as it will have a greater impact in terms of the applied load balancing mechanism.
\par Furthermore, in Fig. 2, the time instances of upward and downward changes are shown in the form of a boxplot. It is intuitive that upward changes occur earlier than downward changes. Moreover, Fig. 2 demonstrates that the multitude of upward changes is greater than the respective of downward changes, indicating that decreases in popularity are sharper than increases. In particular, we estimated that out of the total number of changes,  67$\%$ are upward. 
\begin{figure}
\centering
\begin{subfigure}[b]{0.25\textwidth}           \includegraphics[width=\textwidth]{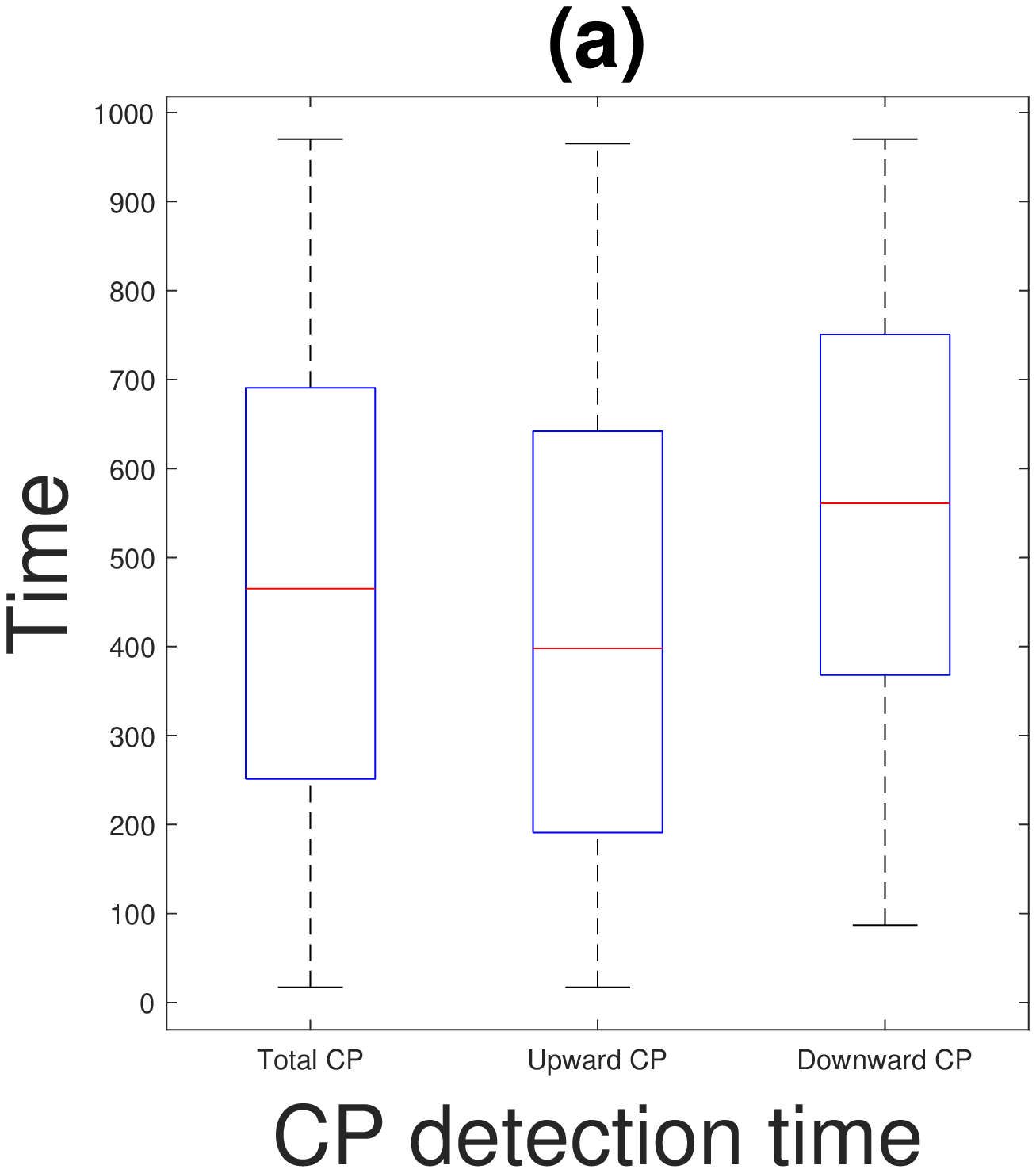}
\end{subfigure}%
\begin{subfigure}[b]{0.25\textwidth}
 \includegraphics[width=\textwidth]{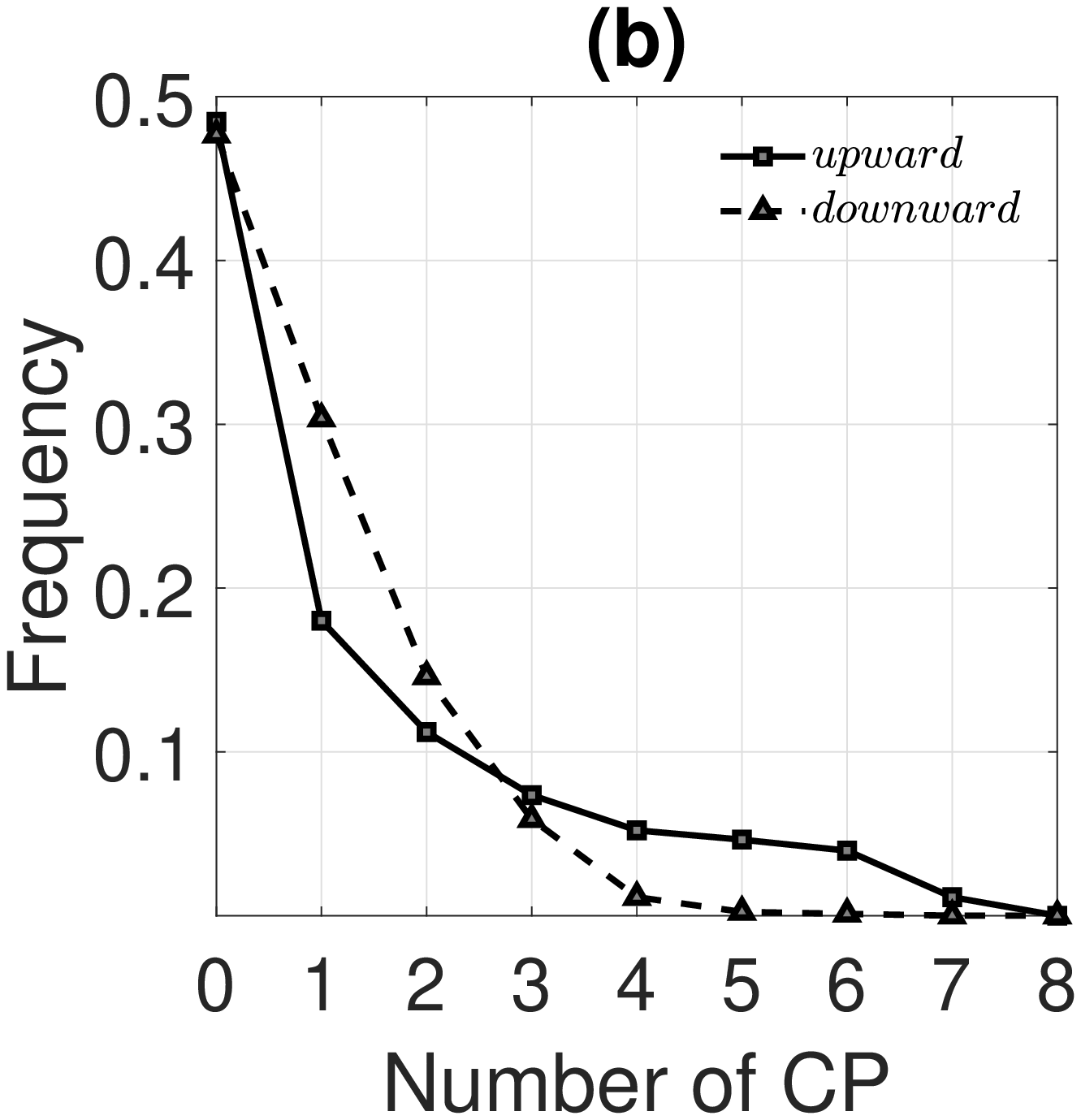}
\end{subfigure}
\caption{Frequency values of the number of upward and downward CPs, per time-series.}
\end{figure}
 
 \begin{figure}
\centering
    \begin{subfigure}[b]{0.25\textwidth}
     \includegraphics[width=\textwidth]{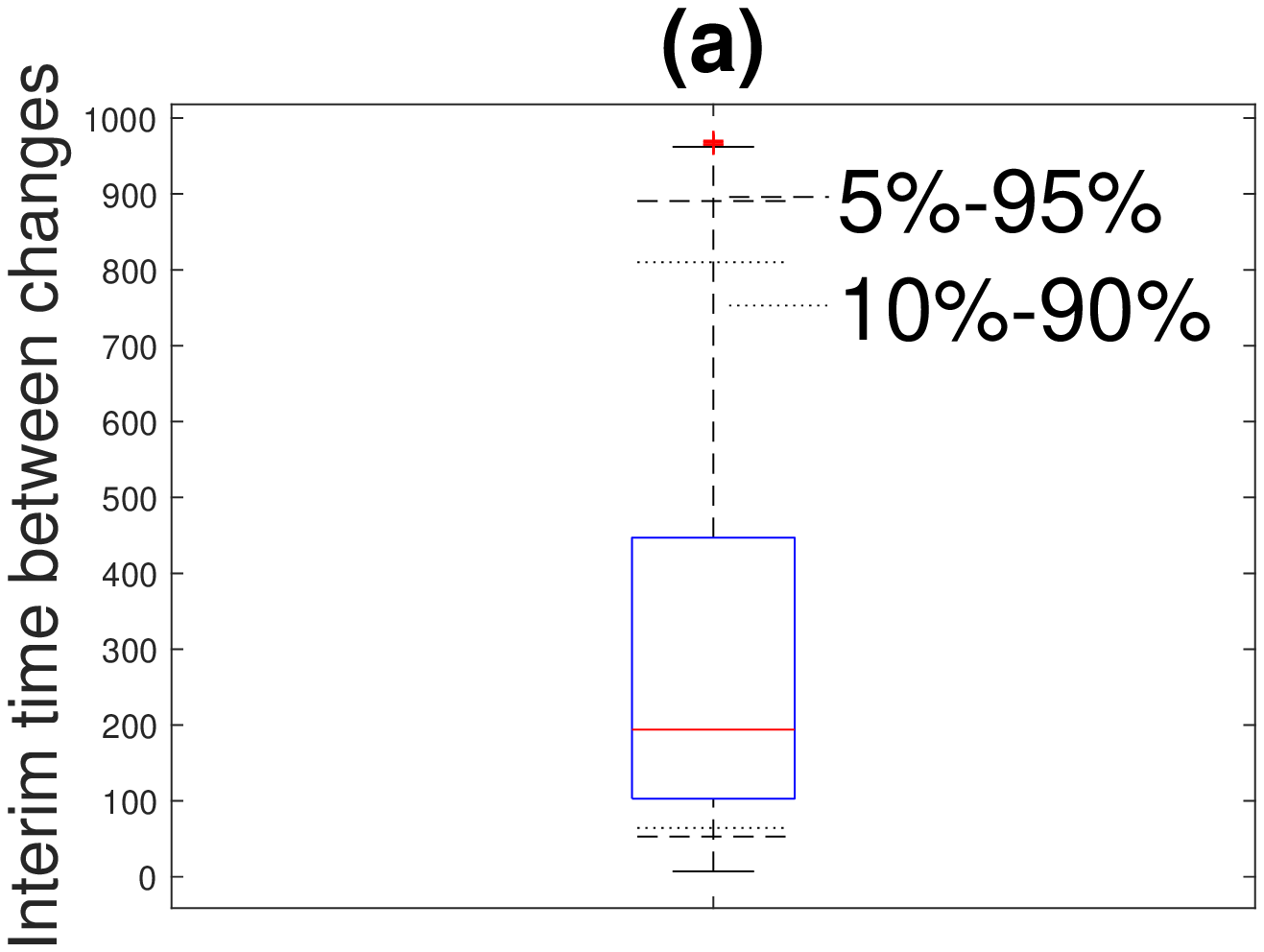}
    \end{subfigure}%
    \begin{subfigure}[b]{0.25\textwidth}
      \includegraphics[width=\textwidth]{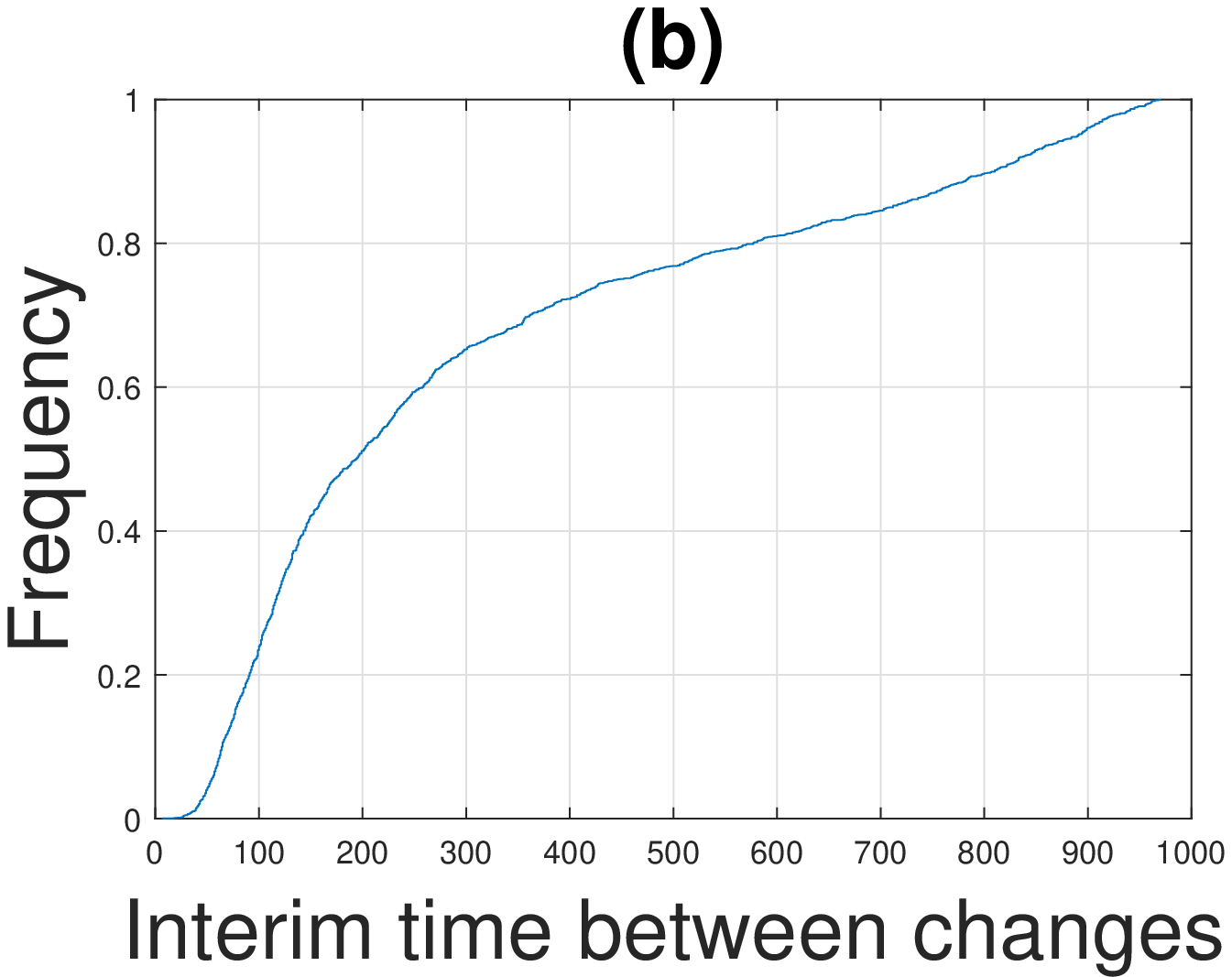}
    \end{subfigure}
\caption{a) Boxplot including the interval $(5\%-95\%)$ (dashed line) and $(10\%-90\%)$ interval (dotted line), b) Cumulative frequency for the interim time of consecutive CPs.}
\end{figure}
 
Finally, we analyze the interim time between consecutive CPs. The results presented in Fig. 3 illustrate the existence of a sufficiently large gap between consecutive potential changes. $90\%$ of the intervals corresponding to consecutive CPs exceed $70$ time instances and only $5\%$ of them are shorter than $50$ time instances, ensuring that a sufficiently large training window can be applied. The results depicted in Fig. 3 allow adjusting parameters of the on-line phase, in particular the minimum time interval between consecutive changes, denoted by the parameter $d$.   

\subsection{Evaluation of the RCPD Algorithm}   
\begin{figure}[t]\label{DTW}
	\centering
 	\includegraphics[width=3.6in]{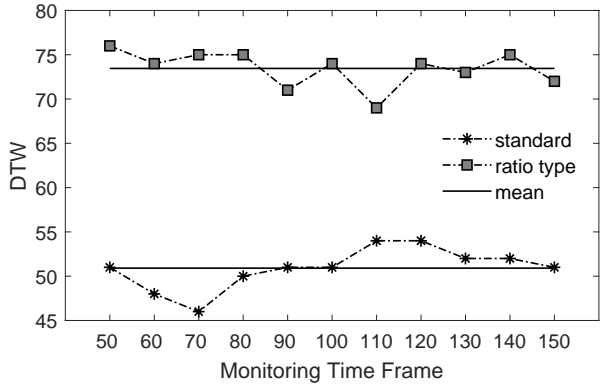}
	\caption{DTW distances for the two on-line detection schemes.}
	\end{figure}

	\par In the previous subsection we have evaluated the performance of the off-line algorithm and demonstrated its efficiency as well as how it is employed in determining parameters of the on-line phase, such as the interval assuming no change $d$ and the threshold parameter of $TI_{f}$ $h$.  
    
    We further employ the off-line algorithm as a benchmark against which the performance of the RCPD algorithm will be evaluated. We note that the off-line analysis provides the \textit{best possible statistical detection} of the actual mean changes, as off-line algorithms operate retrospectively over the entirety of each of the time-series. Thus, in absence of a priori knowledge of the actual CPs in the real data (as opposed to the synthetic data in which the CPs were controlled), we evaluate the performance of the RCPD procedure by measuring the ``similarity'' of its outputs (detected CPs, instances of detection and trends) to the corresponding outputs of the off-line version. 
    
    \par As the number of detected CPs and / or their exact positions are likely to differ at the output of the retrospective (off-line) and of the RCPD algorithm, in order to obtain a measure of their similarity, we estimate their dynamic time warping (DTW) distance.  The DTW is a dynamic programming tool that measures distances between asynchronous sequences and is widely used by the speech processing community \cite{dtw}.  
    
    The results are presented in Fig. 4, where the estimated DTW distances are depicted for several values of the monitoring window length $l\in [40,150]$, to investigate the consistency of parameter $l$ over different values. In the RCPD algorithm we use $d=50$ (minimum distance between two changes) and have set the sensitivity parameter to $\gamma=0.25$. The estimated mean DTW distance for the standard CUSUM is $52$ and for the ratio-type CUSUM is $73$. For comparison purposes, we note that the corresponding DTW distance over the synthetic data is $20$ for medium / large changes, while the true CP detections are around $95\%$. As a result, we can infer, that the outputs of the on-line algorithm, using the standard CUSUM, are ``very close'' to the outputs of the benchmark off-line algorithm. In agreement with our observations over the synthetic data, the DTW distance using the ratio-type CUSUM is clearly larger. 
    \begin{table}[!t]
\renewcommand{\arraystretch}{1.2}
\setlength{\tabcolsep}{12pt}
\centering
\caption{Empirical percentiles of mean values change rate.}
\begin{tabular}{c|cccc}
\hline
\hline
 & \multicolumn{4}{c}{Percentiles Threshold}\\
\hline
\multicolumn{1}{c|}{ } & \multicolumn{1}{c}{10\%} & \multicolumn{1}{c}{15\%} & \multicolumn{1}{c}{25\%} & \multicolumn{1}{c}{50\%}\\
\hline
\multirow{1}{*}{Standard} & 9\% & 13.1\% & 20.8\% & 42.21\%\\
\hline
\multirow{1}{*}{Ratio type}& 9.5\% & 14.82\% & 28.22\% & 67.40\%\\
\hline
\end{tabular}\label{ta:percentiles thershold}
\end{table} 

We also study the magnitude of the detected CPs. We define as the CP magnitude the percentage-wise change in the mean values before and after the CP. We group the measured magnitudes for all change points using the four percentile threshold values $10\%$, $15\%$, $25\%$ and $50\%$, i.e., reflecting the frequency of magnitudes exceeding the respective thresholds. The results are summarized in Table \ref{ta:percentiles thershold}. According to our results, both the standard and ratio type CUSUM algorithms detect the most significant changes in the content popularity. Moreover, 
ratio-type CUSUM detects, in general, CPs with the largest magnitude of change, in agreement with synthetic data results.    

    

        \begin{figure*}
\renewcommand{\arraystretch}{1.5}
    \centering
    \begin{subfigure}[b]{0.25\textwidth}
    \includegraphics[width=\textwidth]{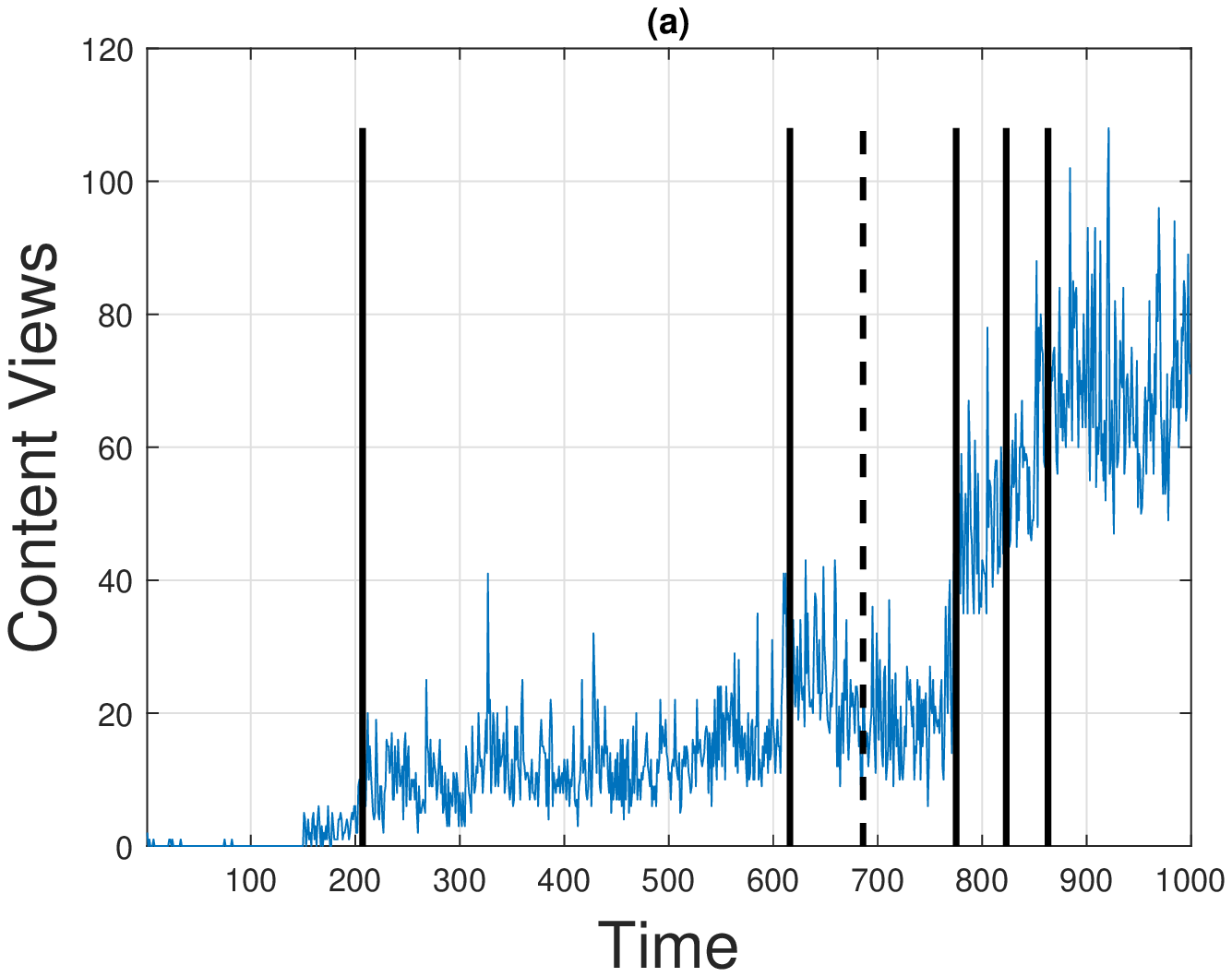}
    \end{subfigure}%
        \begin{subfigure}[b]{0.25\textwidth}
            \includegraphics[width=\textwidth]{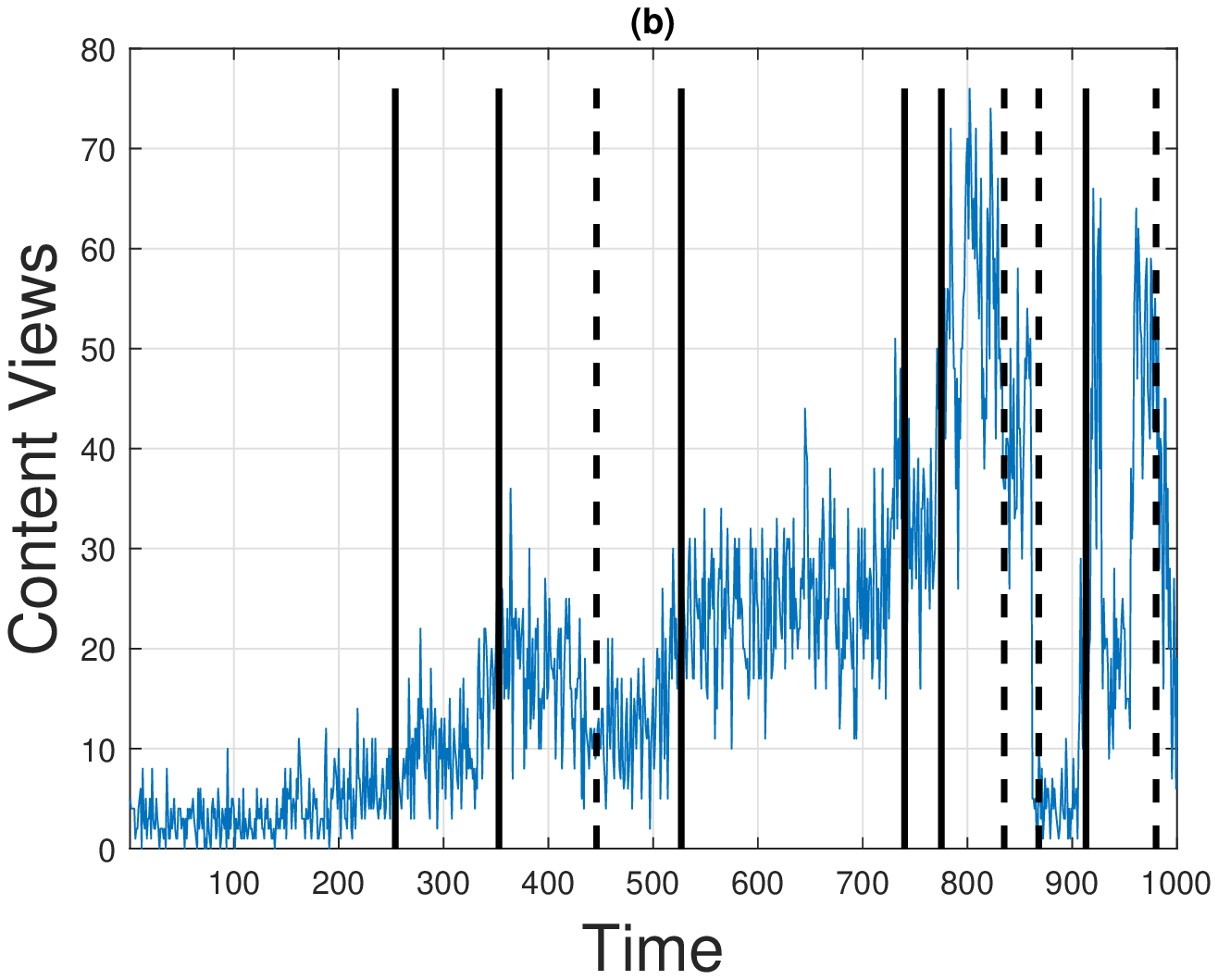}
    \end{subfigure}%
    \begin{subfigure}[b]{0.25\textwidth}
            \includegraphics[width=\textwidth]{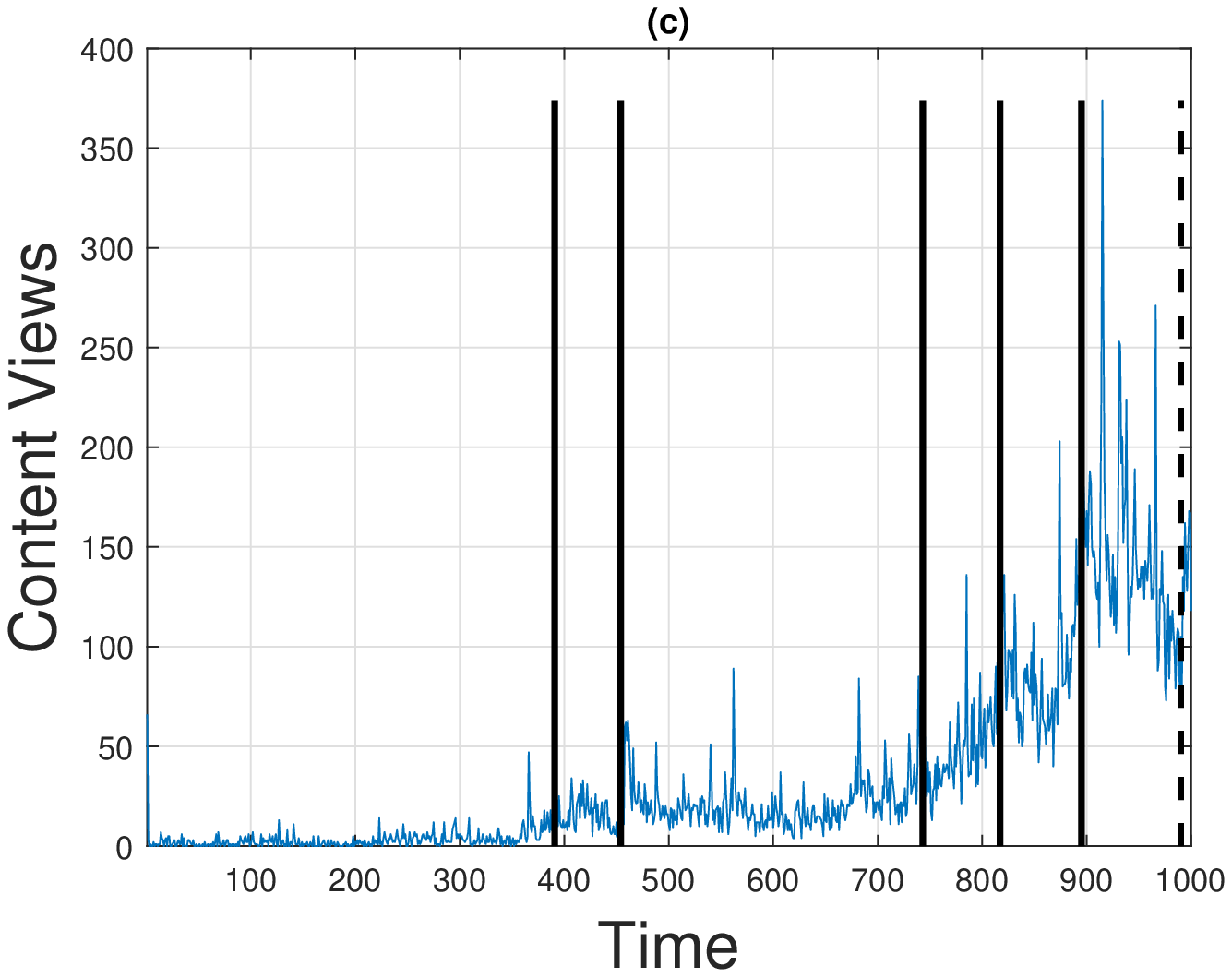}
    \end{subfigure}%
        \begin{subfigure}[b]{0.25\textwidth}
            \includegraphics[width=\textwidth]{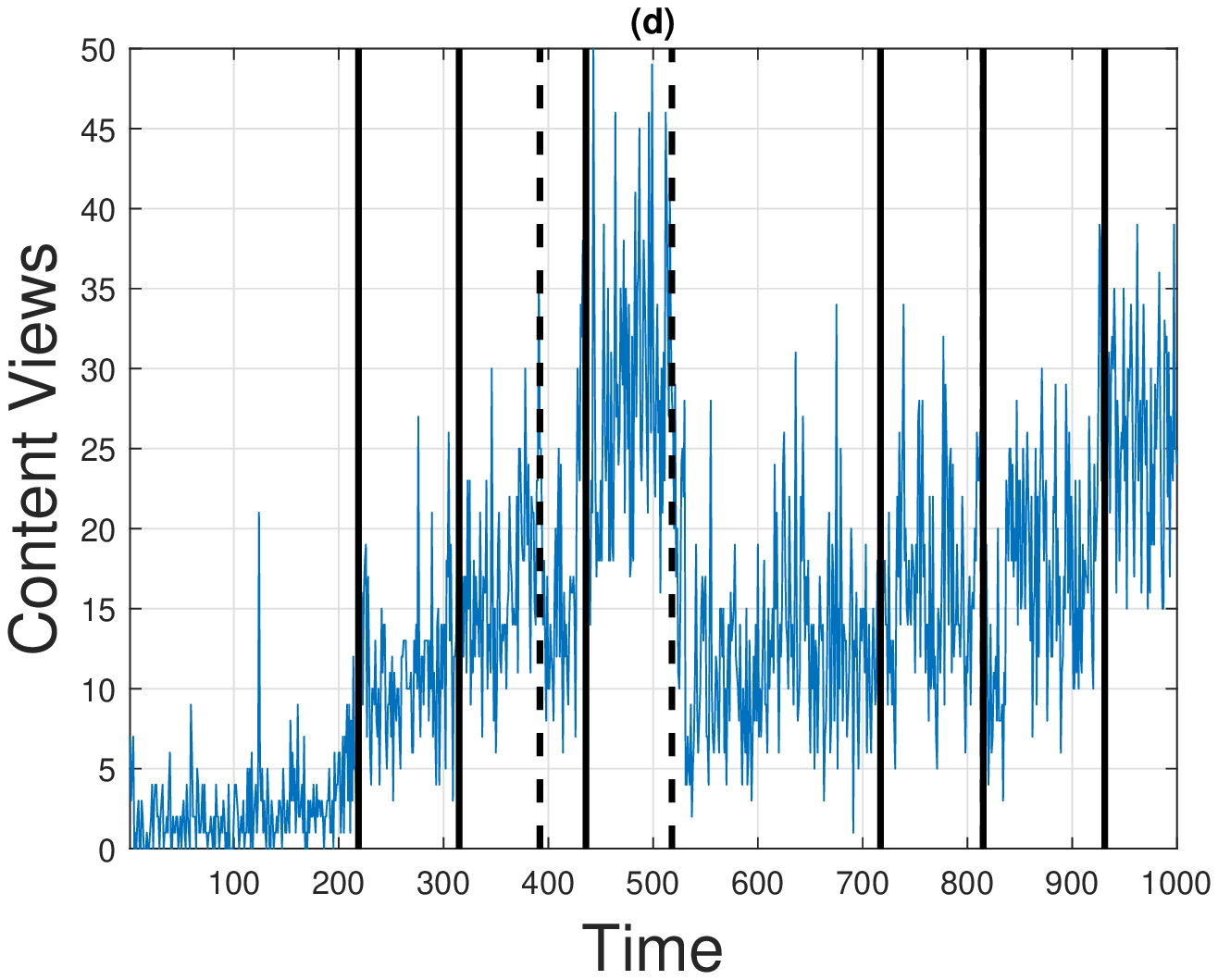}
    \end{subfigure}
    \begin{subfigure}[b]{0.25\textwidth}
           \includegraphics[width=\textwidth]{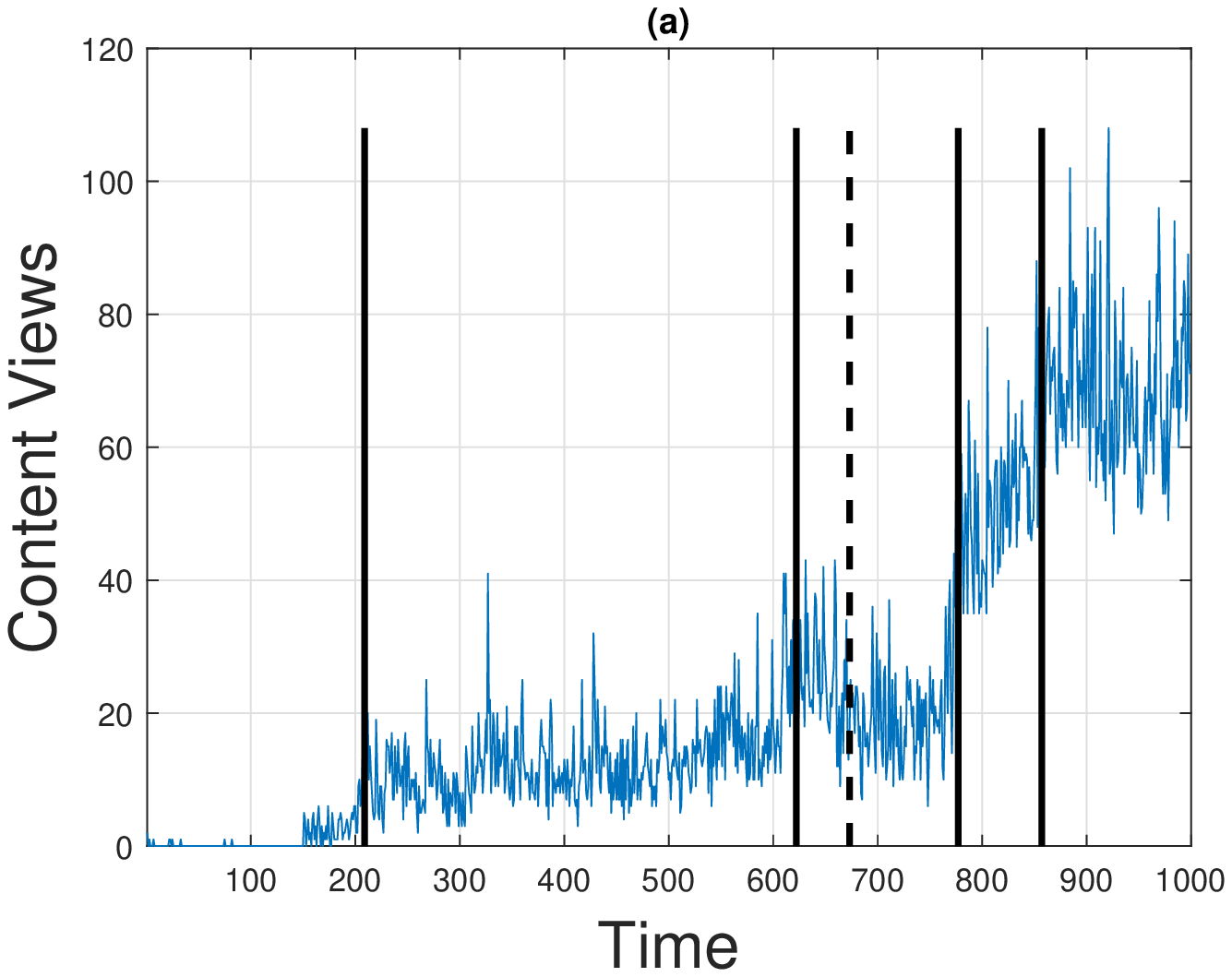}
    \end{subfigure}%
    \begin{subfigure}[b]{0.25\textwidth}
           \includegraphics[width=\textwidth]{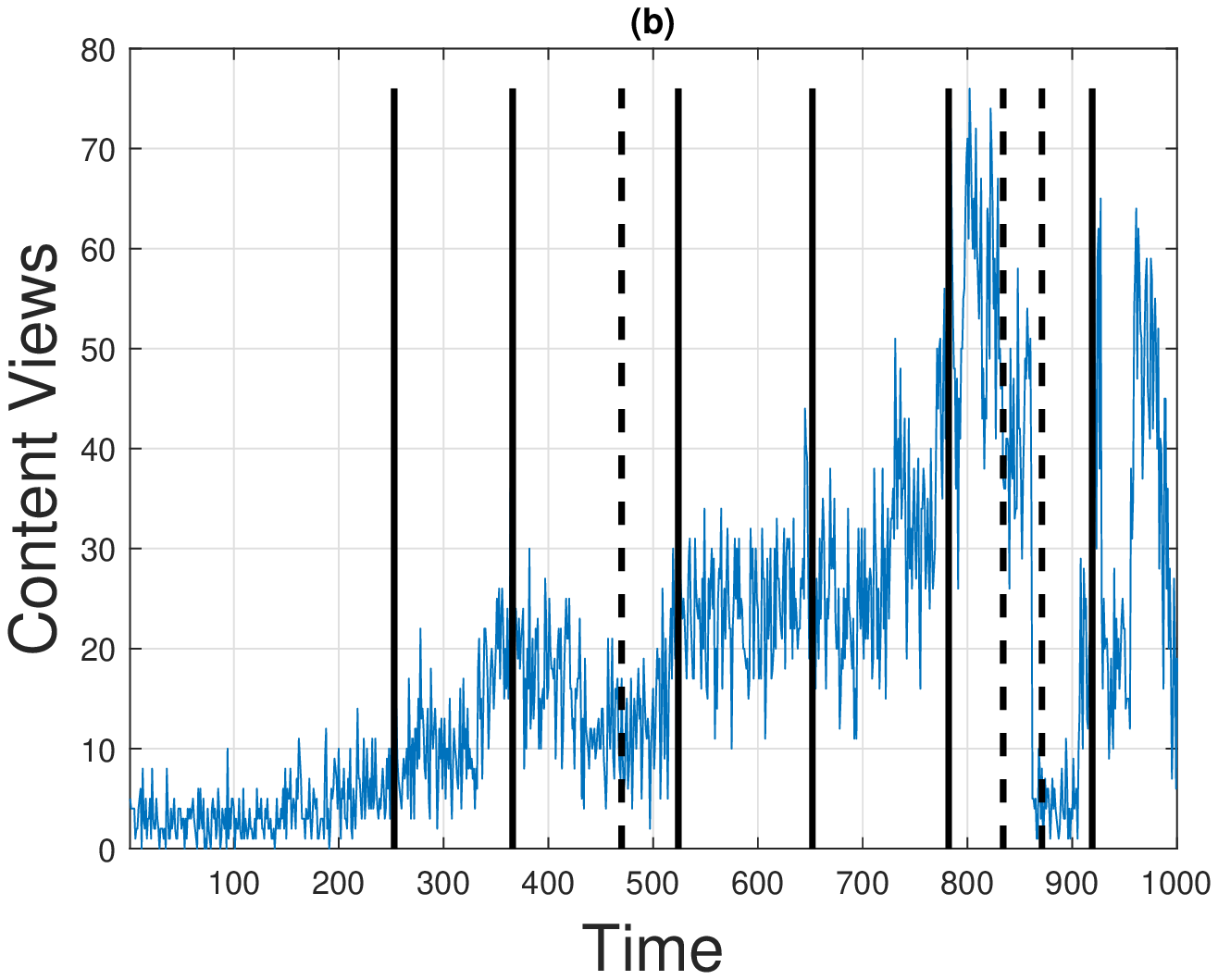}
    \end{subfigure}%
    \begin{subfigure}[b]{0.25\textwidth}
           \includegraphics[width=\textwidth]{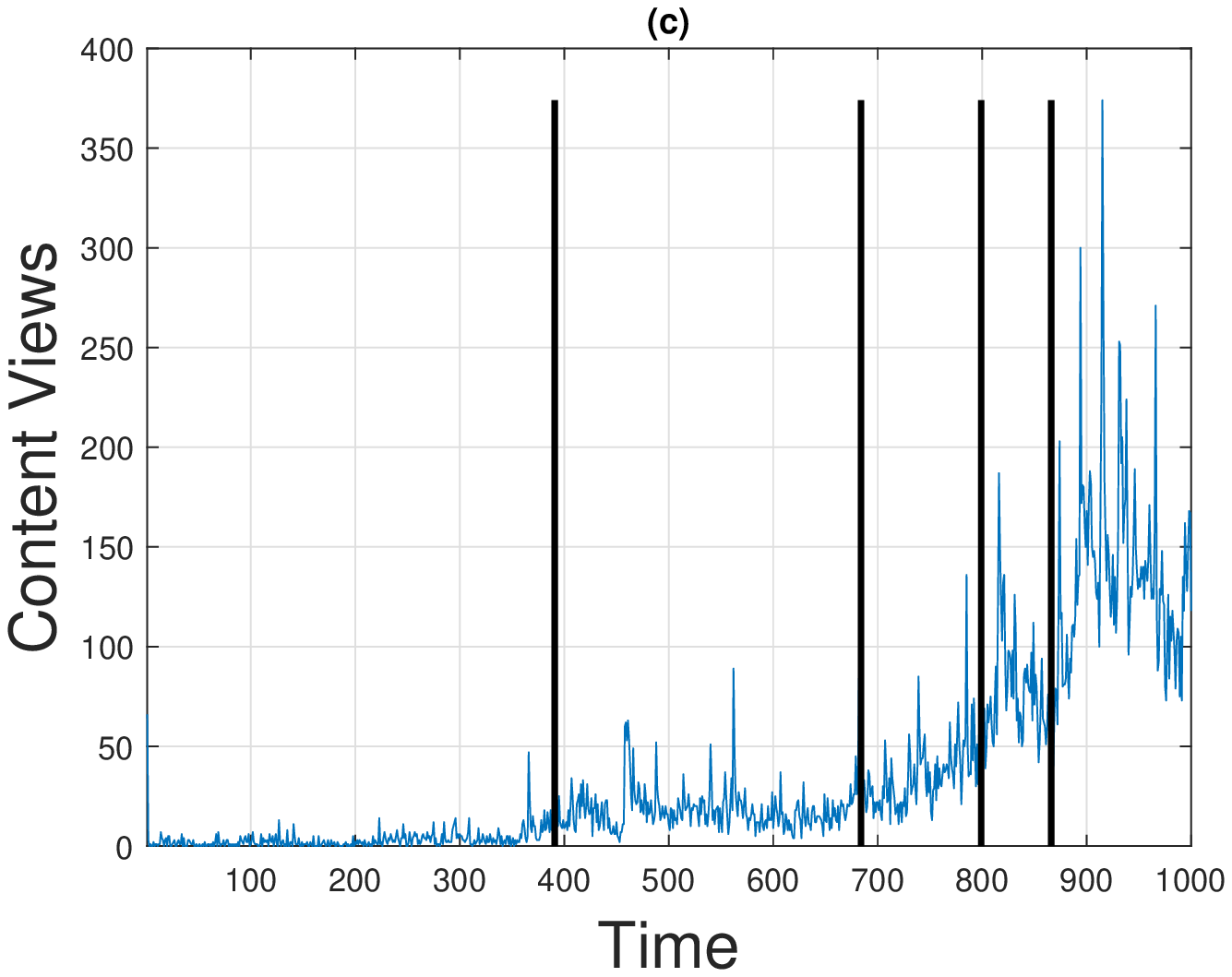}
    \end{subfigure}%
    \begin{subfigure}[b]{0.25\textwidth}
           \includegraphics[width=\textwidth]{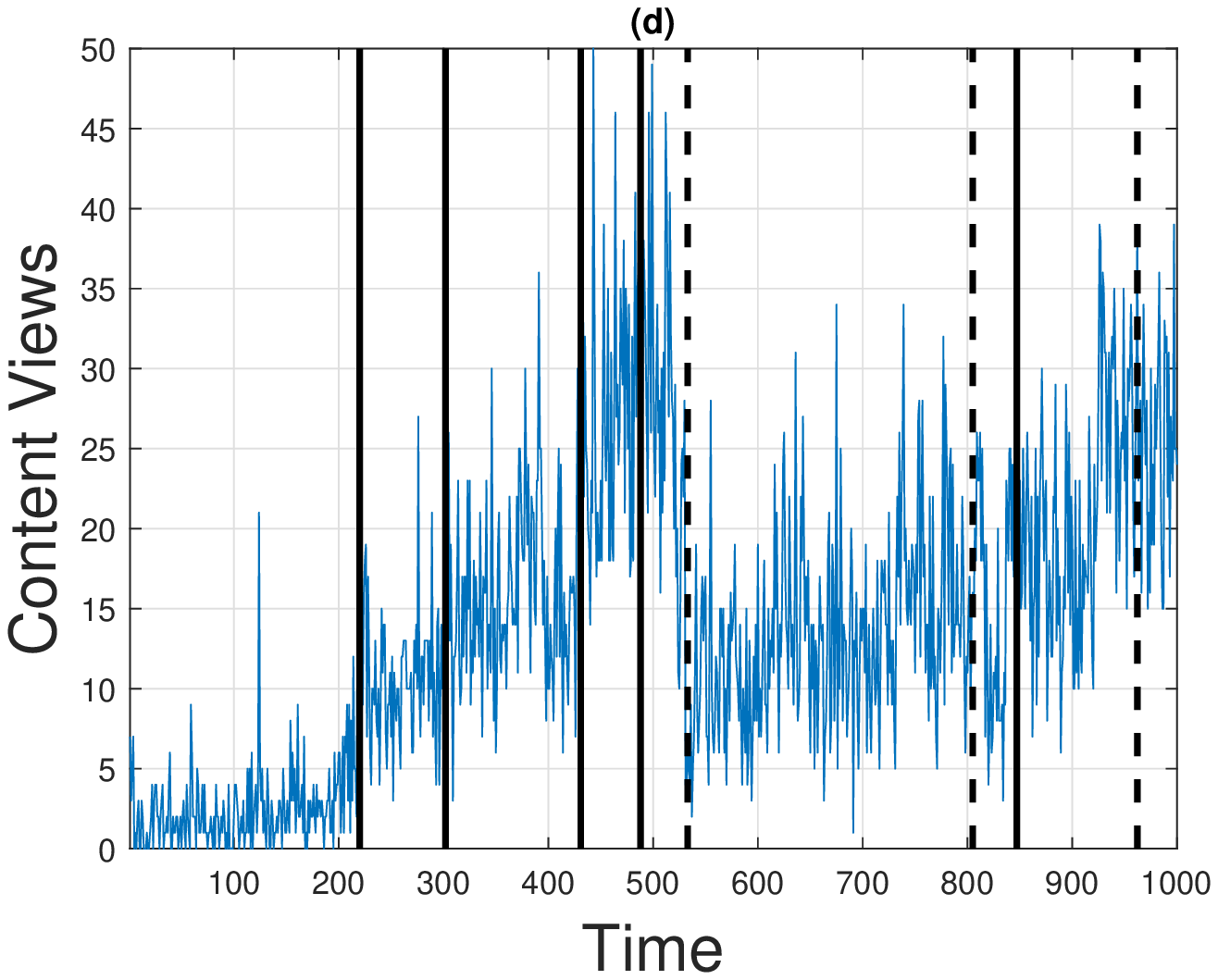}
    \end{subfigure}

    \caption{Outputs of the RCPD algorithm; using standard CUSUM (upper row) and ratio type CUSUM (lower row) for four different time-series. Solid and dashed lines depict an upward and a downward change, respectively.}
\end{figure*}
  
  \begin{figure}\label{load balancing}
\centering
\includegraphics[width=3in]{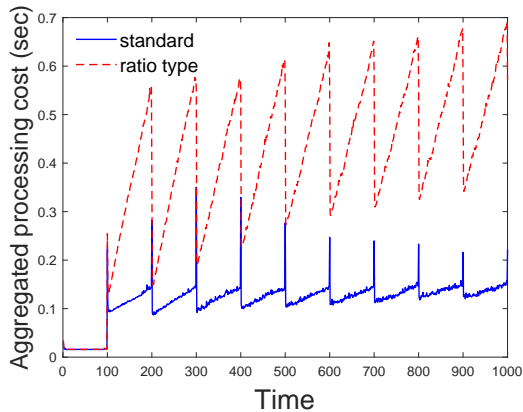}
\caption{The aggregated overall processing cost, per time-instance, of the RCPD algorithm over $882$ time-series.}\label{fig:scalability}
\end{figure}
  
\par Additionally, for illustration purposes, we depict the RCPD algorithm's outputs for four different time-series. We set the beginning of the monitoring period at $m_{s}=200$ and monitoring horizon $l=50$, the on-line parameter $g=0.25$ and the significance level to $a=0.05$. The corresponding results are depicted in Fig. 5, showing the estimated CPs by applying the standard CUSUM and the ratio type CUSUM procedures, respectively. In both cases, the estimated changes correspond to the real content popularity changes; visual inspection suggests that the performance of the standard CUSUM is more reasonable (e.g., Fig. 5d). The RCPD, as it is illustrated in Fig. 5b seems to be adaptable to ``fast'' changes; without getting ``confused'' by random peaks in the time-series, such as those in Fig. 5a or in Fig. 5c.

\subsection{Time Dependencies of Piecewise time-series}
\begin{table}[t]
\renewcommand{\arraystretch}{1.4}
\setlength{\tabcolsep}{10pt}
\centering
\caption{Percentages of time-series with Time Dependencies Exceeding $t$ Samples}
\label{tab:auto_corr}
\begin{tabular}{cccccc}
\hline
\hline
$t$ & $\geq1$ & $\geq5$ & $\geq15$ & $\geq30$ & $\geq50$\\
\hline
\multirow{1}{*}{piecewise} & 0.93 & 0.57 & 0.23 & 0.05 & 0.04\\
\hline
\end{tabular}
\end{table}  

We also measure the autocorrelation function of the piecewise - divided by the detected CPs - time-series. Results are tabulated in 
Table \ref{tab:auto_corr} and verify the short dependence structure of the dataset; significant lags in time dependencies higher than $30$ instances can be found in less than $5\%$ of the time-series. Furthermore, the fact that the ACF of the piecewise time-series drops to zero quickly indicates that the detected CPs split the time-series into stationary segments, which, additionally, confirms indirectly the accuracy of the off-line CP estimations over the changes in the real data. 

\subsection{Computational Complexity and Scalability}
Finally, we present a MATLAB $\circledR$ implementation of the overall algorithm with a large number of time-series ($882$ in this experiment) to quantify its performance in terms of processing cost. The computational time is measured on a Lenovo IdeaPad 510-15IKB laptop, with an Intel Core i7-7500U @ $2.70$ GHz processor and $12$ GB RAM. In Fig. \ref{fig:scalability}, we show the aggregate processing cost per time instance for the two on-line methods and the total number of time-series. For the first $100$ time instances, the algorithm collects the initial data, since it bootstraps. The peaks indicate the off-line part of the algorithm, which is more processing demanding mainly due to the segmentation algorithms running in parallel. The on-line part in the standard on-line algorithm indicates a linear complexity, since it is based on (\ref{eq:17}), while the equivalent quantity in (\ref{eq:19}) of the ratio-type is more CPU intensive, justifying the comparatively higher processing cost of the latter algorithm. In both cases, the aggregate processing cost is typically much less than a second, which demonstrates the lightweight nature of the proposed scheme. Such results could be further improved with a distributed deployment of scheme replicas since each of the time-series could be processed independently. 

\section{The RCPD Algorithm in a Load Balancing Scenario}

In this Section, we demonstrate our proposal in a real content distribution scenario, balancing the traffic between web clients and content caches with a bespoke DNS-based load-balancer. We implement the RCPD algorithm as a client-server MATLAB $\circledR$ application. The RCPD engine receives periodic content popularity measurements; if a CP is detected, the corresponding upward or downward changes are signalled to the load balancer. The load balancer: (i) distributes the load between the deployed content caches, in a round-robin fashion; (ii) tracks content visits and communicates them to the RCPD engine; and (iii) deploys or removes content caches based on the RCPD outputs. 

We implement the web clients using with the httpperf tool (https://github.com/httperf/httperf). The number of clients at each time instance is based on a real time-series of YouTube content views, illustrated in Fig. 7a. In practice, an experimental run without the RCPD mechanisms uses three content caches constantly and a run with the RCPD mechanism enabled uses initially two and then three, four and five content caches, after each of the three detected change points, respectively. As we show in Fig. 7b, the web clients improve their connectivity times to download the content, while as demonstrated in Fig. 7c the CPU utilization in the servers hosting the content remains almost the same. A relevant experimental platform is presented in \cite{chronis-catania}.

\begin{figure}[t]\label{load balancing}
      \centering
      \begin{subfigure}[b]{0.5\textwidth}
 	\includegraphics[width=9cm, height=4cm]{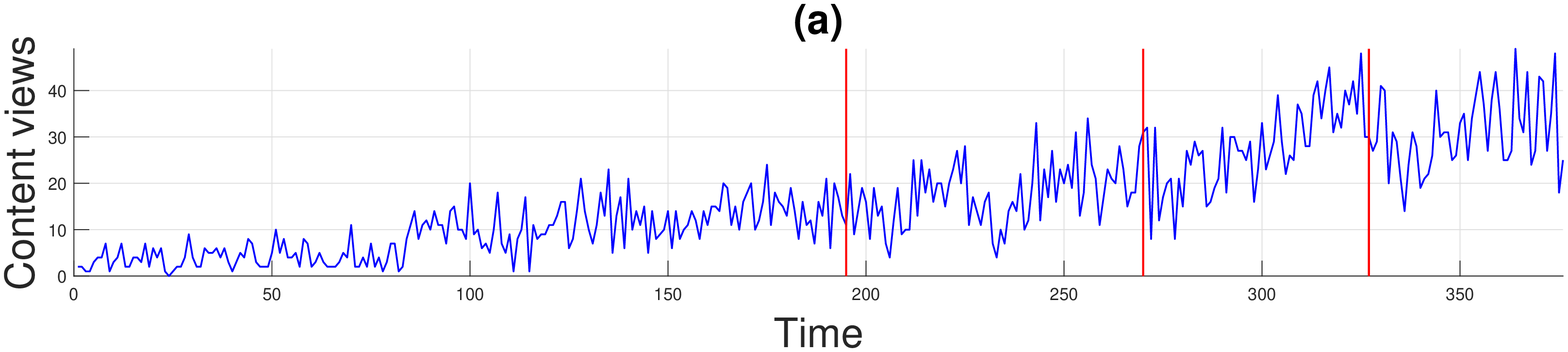}
    \end{subfigure}
    \begin{subfigure}[b]{0.25\textwidth}
 	\includegraphics[width=\textwidth]{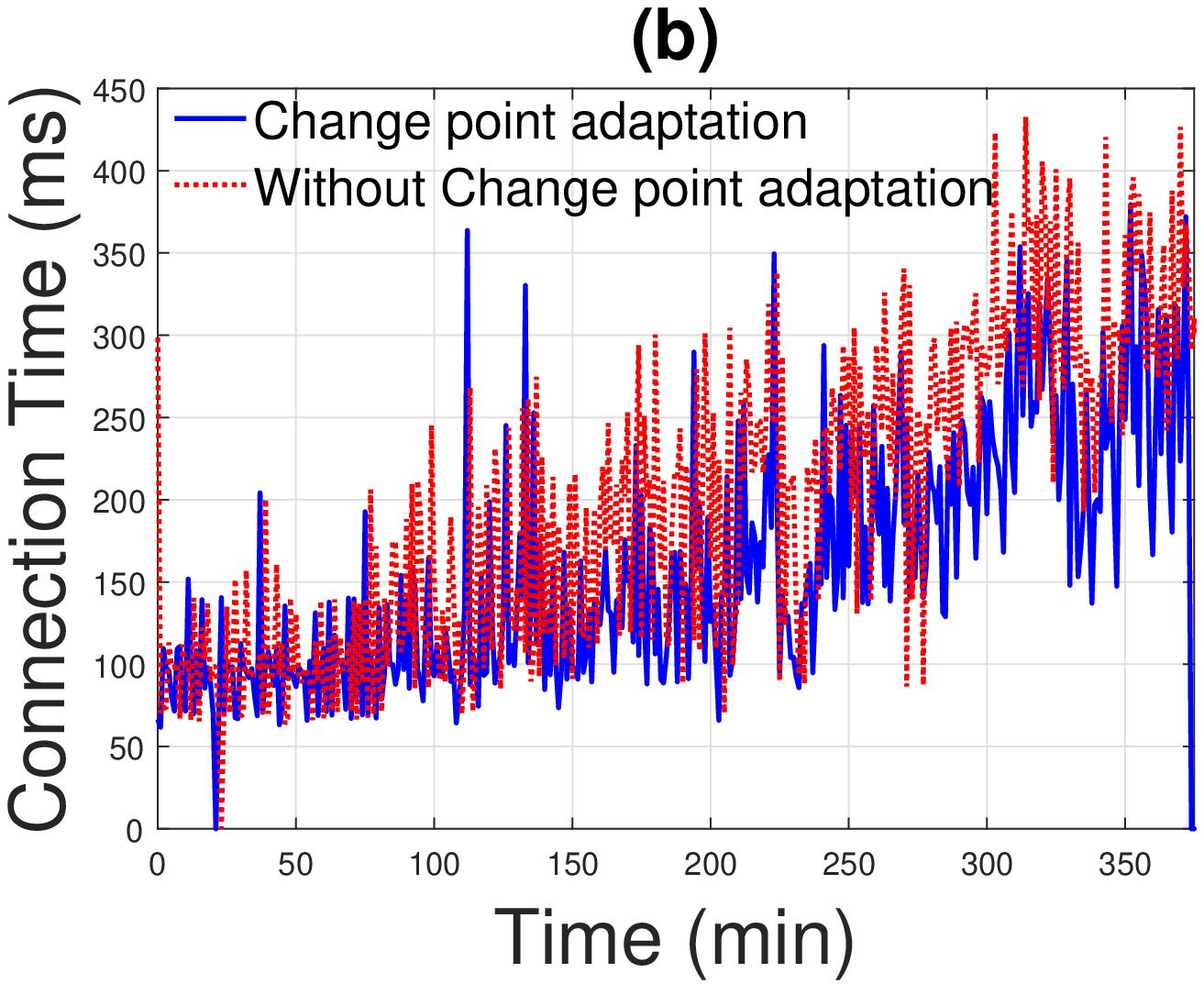}
    \end{subfigure}%
    \begin{subfigure}[b]{0.25\textwidth}
    \centering
 	\includegraphics[width=\textwidth]{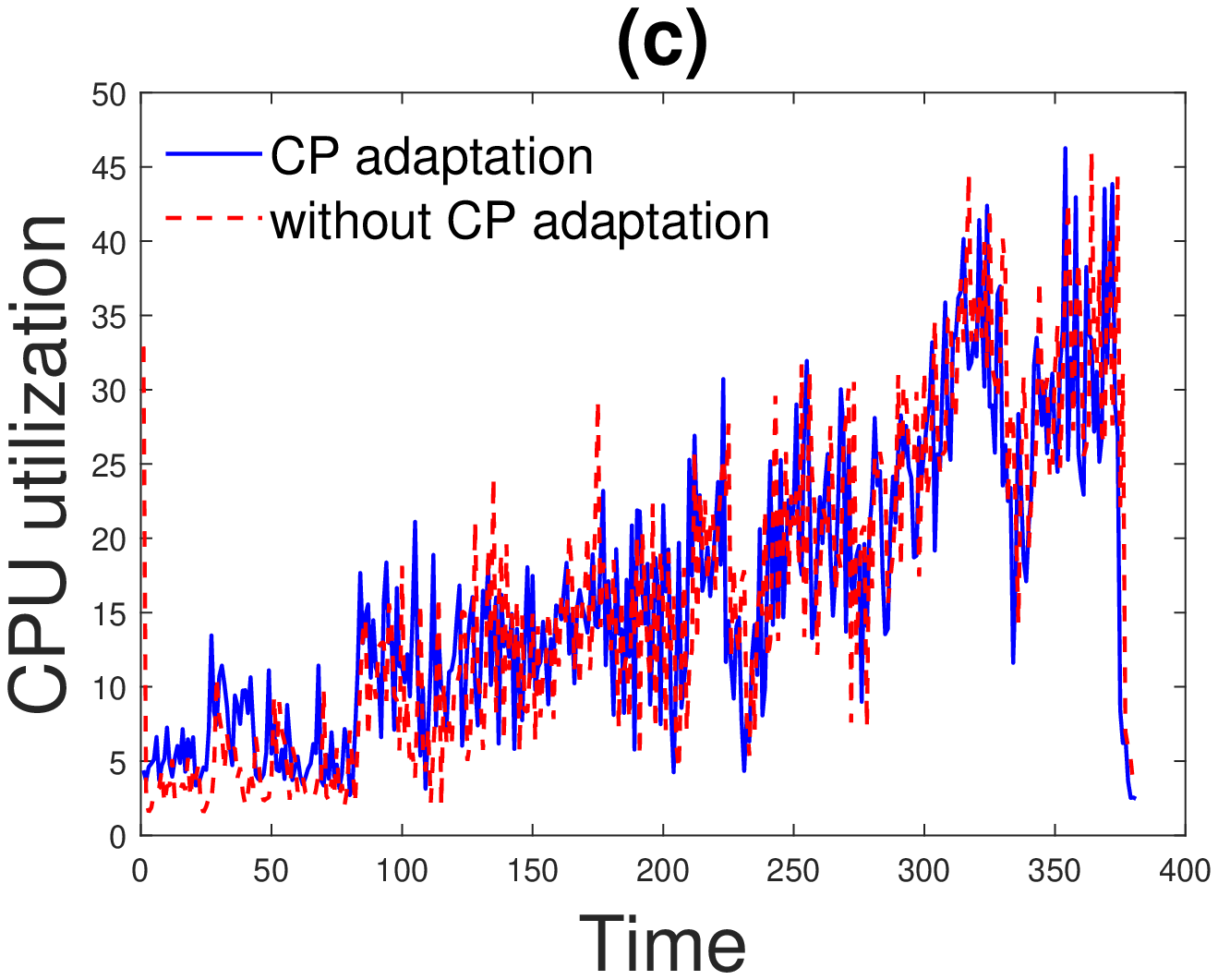}
    \end{subfigure}
	\caption{a) time-series of video content views, red lines depict the detected CPs, b) the connection time with and without RCPD adaptation and c) the equivalent servers’ CPU utilization.}
	\end{figure}
 
\section{Conclusions and Future Work}

\par In this paper, we developed the RCPD, a novel algorithm for the real-time detection of changes in the mean value of content popularity. Approaching the problem statistically, we efficiently combined off-line and on-line non-parametric CUSUM procedures to avoid restrictive assumptions for content popularity behavior and to reduce the overall computational cost. We divided the algorithm in two phases. The first phase is an extended retrospective (off-line) procedure with a modified BS algorithm and is used to adjust on-line parameters, based on historical data of the particular video. The second phase integrates one of two alternative trend indicators to the sequential (on-line) procedure, to reveal the direction of a detected change. We provided extensive simulations, using synthetic and real data, that demonstrated the performance of the proposed algorithm for the successful identification of content popularity changes in real-time. We also demonstrated through experimental measurements that the RCPD's processing cost is almost imperceptible. Finally we provided proof-of-concept by applying the algorithm in a load balancing application, highlighting its efficiency in a realistic setting. 

In future work, we will evaluate the proposed scheme using multi-dimensional time-series to capture more accurately the dynamics of content popularity better (e.g., incorporate additional dimensions with the number of likes, comments, etc.) and in different contexts, such as on the real-time resource utilization of servers. We will also investigate and further extend the algorithm's scalability properties, theoretically and experimentally, i.e., estimate the number of videos that can be analyzed in parallel. Our aspiration is to conduct real large-scale CDN experiments utilizing a distributed architecture with multiple content popularity analyzers, monitoring in real-time clusters of videos at a minimum overall processing cost.   



%



\section*{Acknowledgment}
This work has received funding from the EU's Horizon 2020 research and innovation programme through the 4th open call scheme of the FED4FIRE+ (grant agr. n\textsuperscript{\underline{o}} 732638), the EU-BRA Horizon 2020 NECOS Project (grant agr. n\textsuperscript{\underline{o}} 777067) and the project ``GSRT funding for the years 2016-2017 (Award for the participation in competitive E.U. projects) [Measuring Mobile Broadband Networks in Europe (MONROE, H2020, grant agreement no. $644399$)]", Ministry of Education, Research and Religious Affairs, General Secretariat for Research and Technology (GSRT), Greece. 
%

\bibliographystyle{IEEEtran}
\bibliography{IEEEabrv,reference}

\begin{thebibliography}{10}
\providecommand{\url}[1]{#1}
\csname url@samestyle\endcsname
\providecommand{\newblock}{\relax}
\providecommand{\bibinfo}[2]{#2}
\providecommand{\BIBentrySTDinterwordspacing}{\spaceskip=0pt\relax}
\providecommand{\BIBentryALTinterwordstretchfactor}{4}
\providecommand{\BIBentryALTinterwordspacing}{\spaceskip=\fontdimen2\font plus
\BIBentryALTinterwordstretchfactor\fontdimen3\font minus
  \fontdimen4\font\relax}
\providecommand{\BIBforeignlanguage}[2]{{%
\expandafter\ifx\csname l@#1\endcsname\relax
\typeout{** WARNING: IEEEtran.bst: No hyphenation pattern has been}%
\typeout{** loaded for the language `#1'. Using the pattern for}%
\typeout{** the default language instead.}%
\else
\language=\csname l@#1\endcsname
\fi
#2}}
\providecommand{\BIBdecl}{\relax}
\BIBdecl

\bibitem{ersi-phd-1}
C.~V. Networking, ``Cisco global cloud index: forecast and methodology,
  2015-2020,'' San Jose, CA, USA, {CISCO}, Tech. Rep., 2017.

\bibitem{openflow}
N.~McKeown, T.~Anderson, H.~Balakrishnan, G.~Parulkar, L.~Peterson, J.~Rexford
  \emph{et~al.}, ``Openflow: enabling innovation in campus networks,''
  \emph{ACM SIGCOMM Comput. Commun. Rev.}, vol.~38, no.~2, pp. 69--74, Mar.
  2008.

\bibitem{necos}
``Necos project: Towards lightweight slicing of cloud federated
  infrastructures,''
  https://intrig.dca.fee.unicamp.br/2017/09/05/necos-2-year-eu-brazil-collaborative-project-starting-in-nov2017/.

\bibitem{chronis-catania}
P.~Valsamas, S.~Skaperas, and L.~Mamatas, ``Elastic content distribution based
  on unikernels and change-point analysis,'' in \emph{Proc. 24th Eur. Wireless
  Conf. (EW)}, Catania, Italy, May 2018, pp. 1--7.

\bibitem{springer-survey}
A.~Tatar, M.~D. De~Amorim, S.~Fdida, and P.~Antoniadis, ``A survey on
  predicting the popularity of web content,'' \emph{J. Internet Services
  Appl.}, vol.~5, no.~1, p.~8, Dec. 2014.

\bibitem{survey-30}
G.~Szabo and B.~A. Huberman, ``Predicting the popularity of online content,''
  \emph{Commun. ACM}, vol.~53, no.~8, pp. 80--88, Aug. 2010.

\bibitem{survey-32}
G.~G{\"u}rsun, M.~Crovella, and I.~Matta, ``Describing and forecasting video
  access patterns,'' in \emph{Proc. IEEE Int. Conf. Comput. Commun. (IEEE
  INFOCOM)}, Shanghai, China, Apr. 2011, pp. 16--20.

\bibitem{cascade}
J.~Cheng, L.~Adamic, P.~A. Dow, J.~M. Kleinberg, and J.~Leskovec, ``Can
  cascades be predicted?'' in \emph{Proc. 23rd Int. Conf. World Wide Web
  (WWW)}, Seoul, Republic of Korea, Apr. 2014, pp. 925--936.

\bibitem{b10}
S.~Fremdt, ``Asymptotic distribution of the delay time in page's sequential
  procedure,'' \emph{J. Statist. Planning Inference}, vol. 145, pp. 74--91,
  Feb. 2014.

\bibitem{b11}
Y.~Hoga, ``Monitoring multivariate time series,'' \emph{J. Multivariate Anal.},
  vol. 155, pp. 105--121, Mar. 2017.

\bibitem{brodsky}
E.~Brodsky and B.~S. Darkhovsky, \emph{Nonparametric methods in change point
  problems}, Dordrecht, The Netherlands: Kluwer, 2013.

\bibitem{dtw}
D.~J. Berndt and J.~Clifford, ``Using dynamic time warping to find patterns in
  time series,'' in \emph{Proc. AAAI Workshop Knowl. Disc. Databases (KDD)},
  vol.~10, no.~16, Seattle, USA, Aug. 1994, pp. 359--370.

\bibitem{dtw2}
R.~J. Kate, ``Using dynamic time warping distances as features for improved
  time series classification,'' \emph{Data Mining Knowledge Discovery},
  vol.~30, pp. 283--312, Mar. 2016.

\bibitem{survey-33}
H.~Pinto, J.~M. Almeida, and M.~A. Gon{\c{c}}alves, ``Using early view patterns
  to predict the popularity of youtube videos,'' in \emph{Proc. 6th ACM Int.
  Conf. Web Search and Data Mining (WSDM)}, Rome, Italy, Feb. 2013, pp.
  365--374.

\bibitem{GLOBECOM}
S.~Skaperas, L.~Mamatas, and A.~Chorti, ``Early video content popularity
  detection with change point analysis,'' in \emph{Proc. IEEE Global Commun.
  Conf. (IEEE GLOBECOM)}, Abu Dhabi, UAE, Dec. 2018, pp. 1--7.

\bibitem{basev}
M.~Basseville, I.~V. Nikiforov \emph{et~al.}, \emph{Detection of abrupt
  changes: theory and application}.\hskip 1em plus 0.5em minus 0.4em\relax
  Prentice Hall Englewood Cliffs, 1993, vol. 104.

\bibitem{chandola}
V.~Chandola, A.~Banerjee, and V.~Kumar, ``Anomaly detection: A survey,''
  \emph{ACM Comput. Surveys (CSUR)}, vol.~41, no.~3, pp. 1--58, Sept. 2009.

\bibitem{tarta13}
A.~G. Tartakovsky, A.~S. Polunchenko, and G.~Sokolov, ``Efficient computer
  network anomaly detection by changepoint detection methods,'' \emph{IEEE J.
  Sel. Topics Signal Process.}, vol.~7, no.~1, pp. 4--11, Feb. 2013.

\bibitem{tarta06}
A.~G. Tartakovsky, B.~L. Rozovskii, R.~B. Blazek, and H.~Kim, ``A novel
  approach to detection of intrusions in computer networks via adaptive
  sequential and batch-sequential change-point detection methods,'' \emph{IEEE
  Trans. Signal Process.}, vol.~54, no.~9, pp. 3372--3382, Sept. 2006.

\bibitem{wang}
H.~Wang, D.~Zhang, and K.~G. Shin, ``Change-point monitoring for the detection
  of dos attacks,'' \emph{IEEE Trans. Depend. Sec. Comput.}, vol.~1, no.~4, pp.
  193--208, Oct.-Dec. 2004.

\bibitem{thatte}
G.~Thatte, U.~Mitra, and J.~Heidemann, ``Parametric methods for anomaly
  detection in aggregate traffic,'' \emph{IEEE/ACM Trans. Netw. (TON)},
  vol.~19, no.~2, pp. 512--525, Apr. 2011.

\bibitem{soule}
A.~Soule, K.~Salamatian, and N.~Taft, ``Combining filtering and statistical
  methods for anomaly detection,'' in \emph{Proc. 5th ACM SIGCOMM Conf.
  Internet Measurement}, New York, NY, USA, Oct. 2005, pp. 1--14.

\bibitem{noda}
I.~Nevat, D.~M. Divakaran, S.~G. Nagarajan, P.~Zhang, L.~Su, L.~L. Ko, and
  V.~L. Thing, ``Anomaly detection and attribution in networks with temporally
  correlated traffic,'' \emph{IEEE/ACM Trans. Netw. (TON)}, vol.~26, no.~1, pp.
  131--144, Feb. 2018.

\bibitem{cran}
Y.~Jiang, M.~Ma, M.~Bennis, F.~Zheng, and X.~You, ``A novel caching policy with
  content popularity prediction and user preference learning in fog-ran,'' in
  \emph{Proc. IEEE Global Commun. Conf. (IEEE GLOBECOM) Workshops}, 2017, pp.
  1--6.

\bibitem{phd-31}
X.~Zhou and C.-Z. Xu, ``Optimal video replication and placement on a cluster of
  video-on-demand servers,'' in \emph{in Proc. Int. Conf. Parallel Process.
  (ICPP)}, Vancouver, Canada, Aug. 2002, pp. 547--555.

\bibitem{phd-38}
W.~Tang, Y.~Fu, L.~Cherkasova, and A.~Vahdat, ``Modeling and generating
  realistic streaming media server workloads,'' \emph{Comput. Netw.}, vol.~51,
  no.~1, pp. 336--356, Jan. 2007.

\bibitem{b1}
A.~Aue and L.~Horv{\'a}th, ``Structural breaks in time series,'' \emph{J. Time
  Series Anal.}, vol.~34, no.~1, pp. 1--16, Jan. 2013.

\bibitem{b2}
D.~W. Andrews, ``Heteroskedasticity and autocorrelation consistent covariance
  matrix estimation,'' \emph{Econometrica: J. Econometric Soc.}, vol.~59, pp.
  817--858, May 1991.

\bibitem{b3}
D.~Wied, ``A nonparametric test for a constant correlation matrix,''
  \emph{Econometric Rev.}, vol.~36, no.~10, pp. 1157--1172, Apr. 2017.

\bibitem{lav}
M.~Lavielle and G.~Teyssiere, ``Adaptive detection of multiple change-points in
  asset price volatility,'' in \emph{Long Memory in Economics}.\hskip 1em plus
  0.5em minus 0.4em\relax Springer, G. Teyssiere and A. Kirkman, Eds. Berlin,
  Germany: Springer--Verlag, 2007, pp. 129--156.

\bibitem{ange}
D.~Angelosante and G.~B. Giannakis, ``Sparse graphical modeling of
  piecewise-stationary time series,'' in \emph{Proc. IEEE Int. Conf. Acoust.,
  Speech and Signal Process (IEEE ICASSP)}, Prague, Czech Republic, May 2011,
  pp. 1960--1963.

\bibitem{b8}
C.~Inclan and G.~C. Tiao, ``Use of cumulative sums of squares for retrospective
  detection of changes of variance,'' \emph{J. Amer. Statist. Assoc.}, vol.~89,
  no. 427, pp. 913--923, Sept. 1994.

\bibitem{b5}
H.~Kai, Q.~Zhengwei, and L.~Bo, ``Network anomaly detection based on
  statistical approach and time series analysis,'' in \emph{Proc. Int. Conf.
  Advanced Inform. Netw. Appl. (WAINA) Workshops}, Bradford, UK, May 2009, pp.
  205--211.

\bibitem{Hassine}
N.~B. Hassine, R.~Milocco, and P.~Minet, ``Arma based popularity prediction for
  caching in content delivery networks,'' in \emph{Proc. Wireless Days}, Porto,
  Portugal, Mar. 2017, pp. 113--120.

\bibitem{g}
D.~Wied and P.~Galeano, ``Monitoring correlation change in a sequence of random
  variables,'' \emph{J Statist. Planning Inference}, vol. 143, no.~1, pp.
  186--196, Jan. 2013.

\bibitem{congas}
M.~Zeni, D.~Miorandi, and F.~De~Pellegrini, ``Youstatanalyzer: a tool for
  analysing the dynamics of youtube content popularity,'' in \emph{Proc. 7th
  Int. Conf. Perform. Eval. Methodol. Tools}, Torino, Italy, Dec. 2013, pp.
  286--289.

\bibitem{clegg}
R.~G. Clegg, ``A practical guide to measuring the hurst parameter,'' \emph{Int.
  J. Simul. Syst. Sci. Technol.}, vol.~7, no.~2, pp. 3--14, Nov. 2006.

\bibitem{bardet}
J.-M. Bardet, G.~Lang, G.~Oppenheim, A.~Philippe, S.~Stoev, and M.~S. Taqqu,
  ``Semi-parametric estimation of the long-range dependence parameter: a
  survey,'' \emph{Theory and applications of long-range dependence}, pp.
  557--577, 2003.

\end{thebibliography}

\end{document}